\documentclass[10pt,journal,compsoc]{IEEEtran}

\usepackage{graphicx}
\usepackage{subfigure}
\usepackage{amsmath}
\usepackage{amsfonts}
\usepackage{amssymb}
\usepackage{booktabs}
\usepackage{tablefootnote}
\usepackage{color}

\ifCLASSOPTIONcompsoc
  \usepackage[nocompress]{cite}
\else
  \usepackage{cite}
\fi
\ifCLASSINFOpdf
\else
\fi
\usepackage{url}
\begin{document}
%
\title{SGDA: Towards 3D Universal Pulmonary Nodule Detection via Slice Grouped Domain Attention}
%
%
%
%

\author{Rui Xu,
        Zhi Liu,
        Yong Luo,
        Han Hu,
        Li Shen,
        Bo Du,~\IEEEmembership{Senior Member,~IEEE},\\
        Kaiming Kuang,
        and~Jiancheng Yang
\IEEEcompsocitemizethanks{
\IEEEcompsocthanksitem R. Xu, Y. Luo, and B. Du are with the School of Computer Science, National Engineering Research Center for Multimedia Software, Institute of Artificial Intelligence and Hubei Key Laboratory of Multimedia and Network Communication Engineering, Wuhan University, and Hubei Luojia Laboratory, Wuhan, China. E-mail: rui.xu@whu.edu.cn, yluo180@gmail.com, dubo@whu.edu.cn. \hfil
\IEEEcompsocthanksitem Z. Liu is with the Department of Computer and Network Engineering, The University of Electro-Communications, 1-5-1, Chofugaoka, Chofu-shi, Tokyo,182-8585 Japan. Email: liu@ieee.org. \hfil
\IEEEcompsocthanksitem H. Hu is with the School of Information and Electronics, Beijing Insititute of Technology, Beijing 100081, China. E-mail: hhu@bit.edu.cn. \hfil
\IEEEcompsocthanksitem L. Shen is with the JD Explore Academy, Beijing, China. E-mail: mathshenli@gmail.com. \hfil
\IEEEcompsocthanksitem K. Kuang is with the Dianei Technology, Shanghai 200120, China. E-mail: kaiming.kuang@dianei-ai.com. \hfil
\IEEEcompsocthanksitem J. Yang is with the Shanghai Jiao Tong University, Shanghai, China, Dianei Technology, Shanghai, China, and EPFL, Lausanne, Switzerland. E-mail: jekyll4168@sjtu.edu.cn. \hfil\break
(Corresponding authors: Yong Luo, Bo Du).
}
}

%
%

\markboth{IEEE/ACM TRANSACTIONS ON COMPUTATIONAL BIOLOGY AND BIOINFORMATICS}%
{Shell \MakeLowercase{\textit{et al.}}: }
\IEEEtitleabstractindextext{%
\begin{abstract}
Lung cancer is the leading cause of cancer death worldwide. The best solution for lung cancer is to diagnose the pulmonary nodules in the early stage, which is usually accomplished with the aid of thoracic computed tomography (CT). As deep learning thrives, convolutional neural networks (CNNs) have been introduced into pulmonary nodule detection to help doctors in this labor-intensive task and demonstrated to be very effective. However, the current pulmonary nodule detection methods are usually domain-specific, and cannot satisfy the requirement of working in diverse real-world scenarios. To address this issue, we propose a slice grouped domain attention (SGDA) module to enhance the generalization capability of the pulmonary nodule detection networks. This attention module works in the axial, coronal, and sagittal directions. In each direction, we divide the input feature into groups, and for each group, we utilize a universal adapter bank to capture the feature subspaces of the domains spanned by all pulmonary nodule datasets. Then the bank outputs are combined from the perspective of domain to modulate the input group. Extensive experiments demonstrate that SGDA enables substantially better multi-domain pulmonary nodule detection performance compared with the state-of-the-art multi-domain learning methods.
\end{abstract}

}

\maketitle

\IEEEdisplaynontitleabstractindextext

%
\IEEEpeerreviewmaketitle

\ifCLASSOPTIONcompsoc
\IEEEraisesectionheading{\section{Introduction}\label{sec:Introduction}}
\else
\section{Introduction}
\label{sec:Introduction}
\fi

%
%
%
%
\IEEEPARstart{L}ung cancer has been the most common cause of cancer death in the world. Prompt diagnosis of the pulmonary nodules and timely treatment can significantly improve lung cancer survival rates. For pulmonary nodule detection, the most effective and widely used tool is the thoracic computed tomography (CT). However, a single CT scan contains hundreds of slices; thus interpreting CT data for pulmonary nodule diagnosis is a massive work to doctors, and computer-aided algorithms have been developed to assist doctors in this laborious task \cite{DBLP:journals/mia/GinnekenAHVDNMSRFCBGHFGBTBCa10}, \cite{leaky_noisy-or19}. In recent years, with the prosperity of deep learning, convolutional neural networks (CNNs) have been introduced in the field of pulmonary nodule detection. Powered by the availability of several pulmonary nodule datasets, such as LUNA16 \cite{DBLP:journals/mia/SetioTBBBCCDFGG17}, tianchi \cite{tianchi}, russia \cite{russia}, PN9 \cite{mei2021sanet}, etc, CNN based methods have achieved great success, and become the mainstream approach for pulmonary nodule detection.

\begin{figure}[!t]
\centering
\setlength{\abovecaptionskip}{-0.001in}
\includegraphics[width=3.2in]{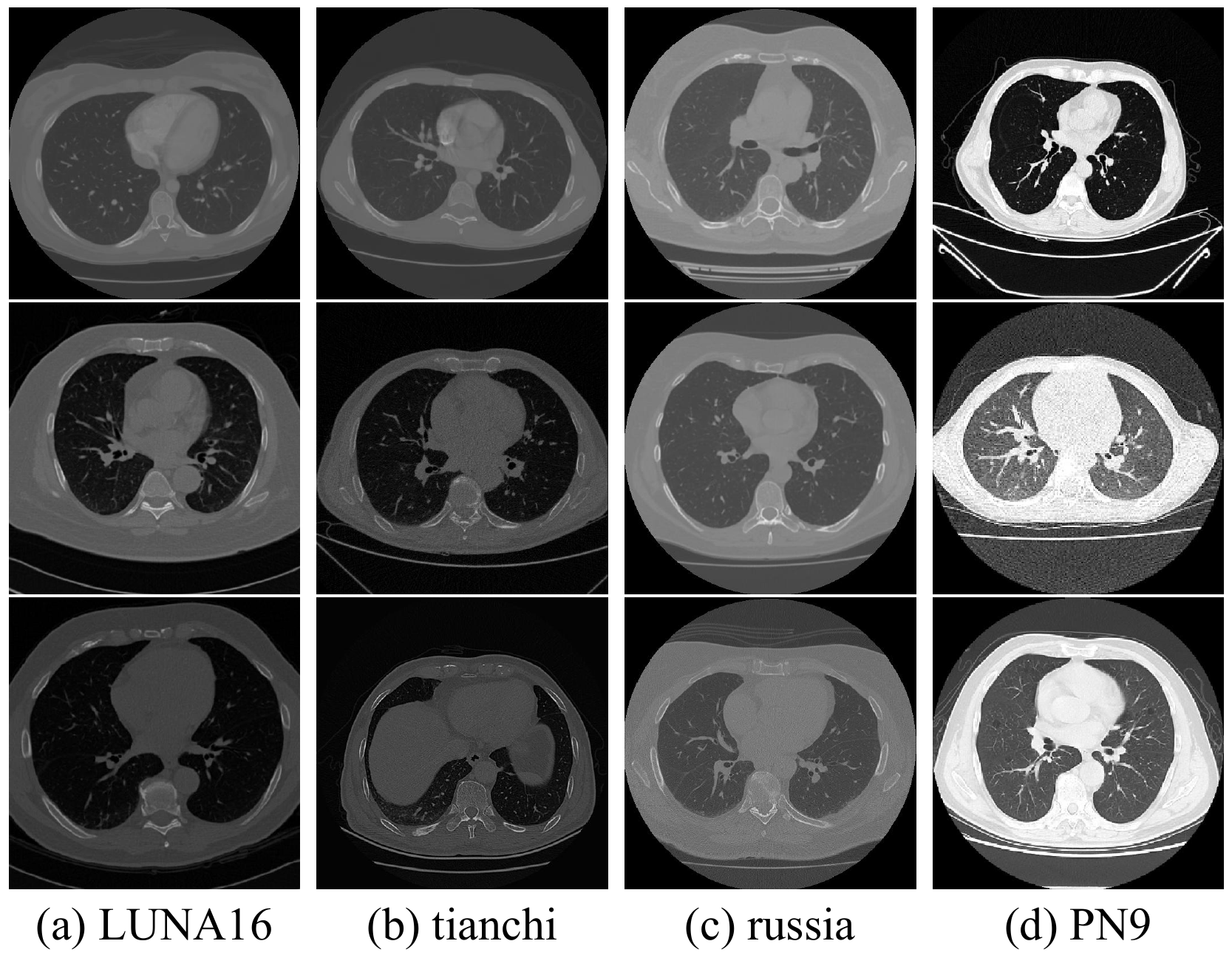}
\caption{Samples of four pulmonary nodule datasets. Each column belongs to one pulmonary nodule dataset as labeled. CT images of different datasets present domain discrepancy, such as illumination, color contrast/saturation, resolution, number of nodules.}
\label{dataset}
\end{figure}

Nonetheless, the existing pulmonary detection methods are usually domain-specific, e.g. trained and tested on the same dataset. They often show performance degradation when applied to other datasets due to the nontrivial domain shift. As shown in Fig. \ref{dataset}, pulmonary nodule datasets can vary in terms of illumination, color contrast/saturation, resolution, etc. The annotation standards of datasets can also be different. For instance, LUNA16 \cite{DBLP:journals/mia/SetioTBBBCCDFGG17} only covers nodules $\geq$ 3mm; russia \cite{russia} just includes binary labels; PN9 \cite{mei2021sanet} not only annotates nodules of all sizes, but also elaborately categorizes them into 9 classes. It is common that high pulmonary detection performance requires a model specially trained for the target dataset. 

However, in most clinical scenarios, multi-center trials have to be conducted. The CT images to be analyzed are not restricted to any one of the domains in Fig. \ref{dataset}. Hence, it is necessary for algorithms capable of detecting nodules from CT scans whichever medical center they are collected from. In natural images, multi-domain learning (MDL) methods are developed to tackle the learning of representations for multiple domains, and which domain the data come from is known as a priori. They often use a combination of parameters shared across domains and parameters specialized for each domain, which are usually known as adapters. Most of them focus on natural image analysis and have achieved significant progress \cite{DBLP:journals/corr/BilenV17},  \cite{DBLP:conf/nips/RebuffiBV17}, \cite{DBLP:conf/cvpr/RebuffiBV18}, \cite{wang2019towards}. In \cite{DBLP:conf/miccai/HuangHYZZ19}, MDL is introduced to medical image segmentation tasks, and a universal architecture is proposed for multi-domain medical image segmentation through parallel channel-wise convolutions, one per domain, followed by one point-wise convolution shared by all domains. A more recent work \cite{DBLP:journals/tmi/YanCZHJTTHXL21} combines MDL with missing annotation mining to develop a universal lesion detection network.
However, existing efforts in this area mostly require prior knowledge of the domain of interest. This is undesirable for autonomous systems in real applications, where determining the domain (data drawn from) is also a nontrivial problem. Therefore, we consider the design of a universal object detection network capable of operating over multiple pulmonary nodule datasets with no need for prior knowledge of the domain of interest.

To achieve this goal, we propose a slice grouped domain attention (SGDA) module for adaptation to different domains, and enhancing the generalization capability of the pulmonary nodule detection networks.
The SGDA module can capture the feature subspaces of the domains spanned by all pulmonary nodule datasets from the axial, coronal, and sagittal directions, and in each direction, we soft-route the projections on these subspaces by group.
Particularly, this domain attention module can be used as a plug-and-play module for existing pulmonary nodule detection networks (such as NoduleNet \cite{10.1007/978-3-030-32226-7_30} and SANet \cite{mei2021sanet}). Taking the widely used NoduleNet \cite{10.1007/978-3-030-32226-7_30} network as an example, this module is added in each 3D residual block \cite{DBLP:conf/cvpr/HeZRS16}.

We summarize our main contributions as follows:
\begin{itemize}
    \item We propose a slice grouped domain attention (SGDA) module, a plug-and-play tool, for existing pulmonary nodule detection networks to enhance their generalization abilities. It mainly contains a universal adapter bank, a domain assignment component, and a three way cross-attention module.
    \item We design a new class of lightweight adapters called slice grouped squeeze-and-excitation (SGSE) adapter to compensate for domain shift.
    \item We introduce domain assignment to achieve domain-aware soft combination of projections on different domains.
    \item We develop three way cross-attention to fuse the modulated feature maps in three directions.
\end{itemize}
To verify the effectiveness of our SGDA, we perform extensive experiments on four pulmonary nodule datasets. Experimental results show that our SGDA outperforms several state-of-the-art multi-domain methods.

The rest of this paper is organized as follows. We summarize the related works of pulmonary nodule detection and multi-domain learning in Section~\ref{sec:Related_Work}. Details of the proposed SGDA method are presented in Section~\ref{sec:Method_SGDA}. Section~\ref{sec:Experiments} includes the experimental results and analysis, and we conclude this paper in Section~\ref{sec:Conclusion}.


\section{Related Work}
\label{sec:Related_Work}

\subsection{Pulmonary Nodule Detection}
Pulmonary nodule detection is usually regarded as an object detection task for CT images, and draws public attention due to its great clinical value. In recent years, CNN based methods have been utilized in various detection tasks \cite{DBLP:conf/iccv/Girshick15}, \cite{DBLP:conf/eccv/LiuAESRFB16}, \cite{DBLP:conf/ijcai/Ye0L020}.
These methods are also introduced in pulmonary nodule detection, and have achieved promising success. Many works are based on 2D CNN. For example, a deconvolutional structure is introduced in Faster RCNN for candidate detection on axial slices \cite{DBLP:conf/miccai/DingLHW17}. In \cite{DBLP:journals/tmi/SetioCLGJRWNSG16}, multi-view ConvNets is proposed for pulmonary nodule detection.
The proposed architecture comprises multiple streams of 2D ConvNets, which take a set of 2D patches from differently oriented planes as input. Then the outputs are combined using a dedicated fusion method to obtain the final results.
More recently, 3D CNN based methods have become the focus of many studies considering the 3D nature of CT images, such as \cite{DBLP:journals/tbe/DouCYQH17}, \cite{deeplung18}, \cite{DBLP:journals/nn/KimYCS19}, \cite{leaky_noisy-or19}, \cite{deepseed20}, \cite{DBLP:journals/tmi/OzdemirRB20}, 
\cite{DBLP:conf/miccai/SongCLHLHCYSZW20}, and \cite{DBLP:journals/mia/LuoSWCCLMZ22}. Specifically, an end-to-end 3D deep CNN called NoduleNet is proposed in \cite{10.1007/978-3-030-32226-7_30} to solve nodule detection, false positive reduction, and nodule segmentation jointly in a multi-task manner. In \cite{mei2021sanet}, a slice-aware network for pulmonary nodule detection termed SANet is developed, which mainly contains a slice grouped non-local module and a false positive reduction module. However, existing pulmonary nodule detectors are usually domain specific, e.g. trained and tested on the same dataset. They may not perform well on other datasets because there exists nontrivial domain shift as shown in Fig.~\ref{dataset}. Likewise, none of these pulmonary nodule detectors could achieve reliable detection performance on diverse datasets/domains of different distributions \cite{xu2021vitae, zhang2022vitaev2}.

\subsection{Multi-Domain Learning/Adaptation}
The concept of multi-domain learning (MDL) is introduced in \cite{DBLP:conf/nips/RebuffiBV17}, which can be regarded as a sub-category of the generic multi-task learning \cite{DBLP:journals/tnn/LuoWT18}, \cite{DBLP:journals/corr/BilenV17}, \cite{DBLP:conf/cvpr/RebuffiBV18}. Different from some general approaches of domain adaptation \cite{mm_WangWLZLTLYL21}, \cite{nn_WangLSWWWZCL22}, \cite{tcsv_WangLWNWL22}, MDL aims to utilize a single model to simultaneously learn multiple diverse visual domains, known as a priori. This can be realized by packing domain-specific parameters in adapters added to the network. The parameters of the resulting network are either shared across domains or domain-specific. For example, in \cite{DBLP:journals/corr/BilenV17}, \cite{DBLP:conf/nips/RebuffiBV17}, and \cite{DBLP:conf/cvpr/RebuffiBV18}, domain-specific normalization layers and domain-specific residual adapters are designed for natural image classification. In \cite{wang2019towards}, a squeeze-and-excitation (SE) \cite{8578843} adapter is proposed for object detection and a domain attention module is further designed for automatic domain assignment. This MDL idea also flourishes in medical imaging analysis. For example, anatomy-specific instance normalization is proposed in \cite{DBLP:conf/miccai/LiuWLZ21} to learn a universal network for under-sampled MRI reconstruction. In \cite{DBLP:conf/miccai/HuangHYZZ19} and \cite{DBLP:conf/miccai/ZhuYXZ21}, separable convolution consisting of domain-specific channel-wise convolution and shared point-wise convolution are explored for medical image segmentation and anatomical landmark detection respectively. In \cite{DBLP:journals/tmi/YanCZHJTTHXL21}, MDL is combined with missing annotation mining to develop a universal lesion detection network for various lesion detection tasks.

Unfortunately, these approaches either do not take full advantage of the 3D nature of medical images, or can not perform automatic inference without prior knowledge of the domains \cite{tmi_ZhuDY20}, \cite{ijcai_ZouZY20}, \cite{miccai_ZhuWYYLL22}. These drawbacks can be remedied by the proposed slice grouped domain attention (SGDA) module, which takes the character of pulmonary nodule detection into consideration, and performs data-driven domain assignments of network activations from the axial, coronal, and sagittal directions.


\begin{figure*}[!t]
\centering
\includegraphics[width=6.5in]{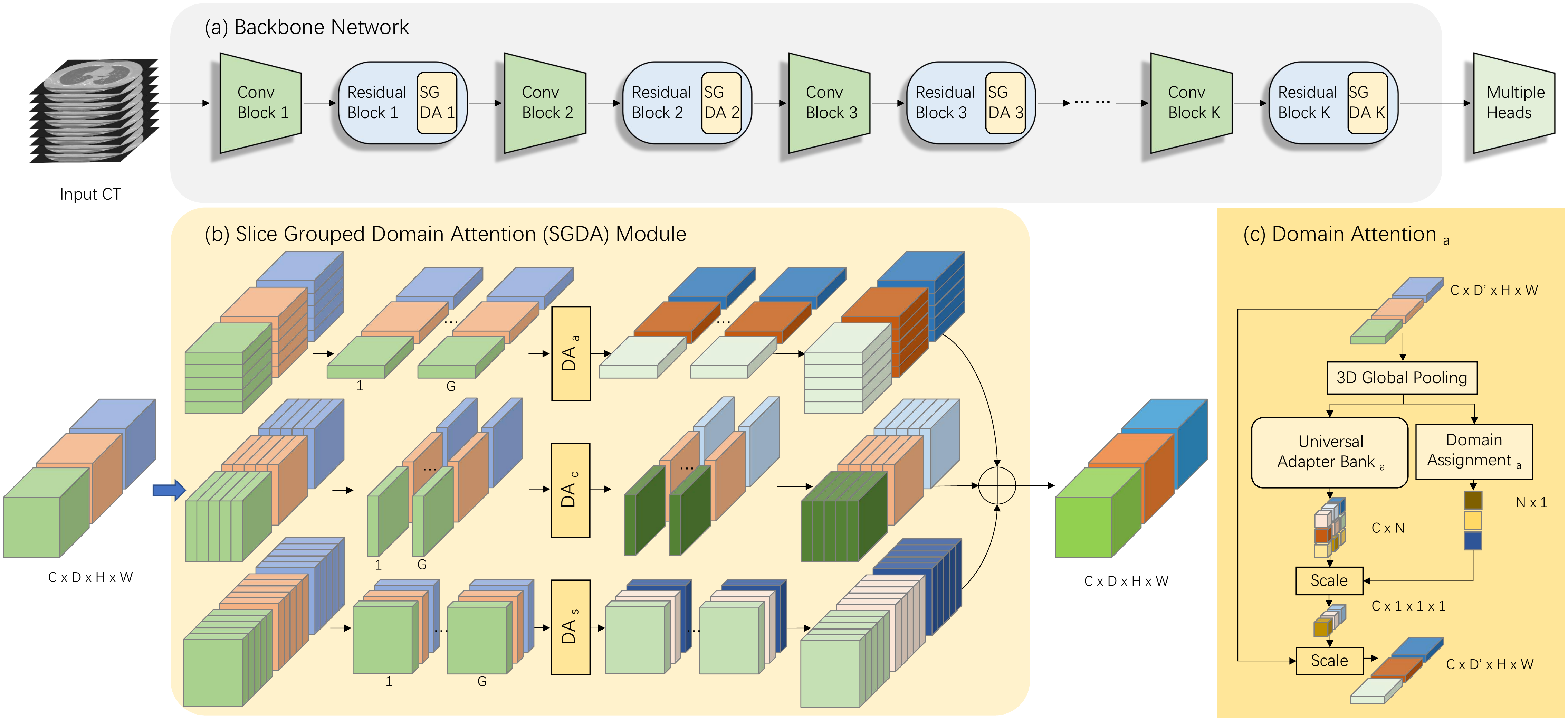}
\caption{(a) Overall flowchart of the universal pulmonary nodule detection network with the proposed Slice Grouped Domain Attention (SGDA) module being plugged in some residual blocks. The backbone of the network is shared across all the datasets, whereas there are multiple detection heads, one for each dataset. (b) The proposed SGDA module. It works in axial, coronal, and sagittal directions to remodulate the input feature map by group, and then fuses the three directional modulated feature maps. (c) Domain attention in one direction consists of a universal adapter bank and a domain assignment component.}
\label{network_frame}
\end{figure*}


\section{Universal Pulmonary Nodule Detection}
\label{sec:Method_SGDA}


In clinical applications, multi-center trials are often required, which involve various datasets/domains. Hence, we aim to design a universal pulmonary nodule detector capable of operating over multiple pulmonary nodule datasets. For a vanilla pulmonary nodule detection task, the goal is to train a detector using one given pulmonary nodule dataset. In this work, we consider a more realistic and complex task, training a universal pulmonary nodule detector on multiple datasets, between which there exist nontrivial domain shifts as shown in Fig.~\ref{dataset}. In other words, the detector needs to be capable of detecting nodules from CT scans no matter which medical center they are collected from, acting like experienced doctors who can diagnose nodules without being affected by domain shift between medical centers.

In order to address the issue that existing pulmonary nodule detectors have little flexibility in dealing with the domain variations in Fig. \ref{dataset}, we propose a slice grouped domain attention (SGDA) module, as illustrated in Fig.~\ref{network_frame}(b).
Particularly, given the 3D feature map of a CT image, we first split the map into slice groups in the axial, coronal, and sagittal directions, so as to explore the inter-dependencies among channels for each group in different directions. Then for each direction, the channel responses of different groups are modulated using the domain attention mechanism, where a new class of light adapters termed slice grouped squeeze-and-excitation (SGSE) adapters are incorporated. This can be seen as a feature based attention. These adapters form as an universal adapter bank, which captures the feature subspaces of the domains spanned by all pulmonary nodule datasets and grouped from the axial, coronal, and sagittal directions. To further achieve automatic inference of domains, a domain assignment component is added to soft-route the universal adapter bank projections by group in three directions. It should be noted that both the projections on domains and domain assignment are data-adaptive and not bind to specific tasks/datasets.
Afterwards, the modulated groups are stacked and the resulting feature maps in the three directions are fused by a three way cross attention module (instead of simple summation). Finally, we incorporate the SGDA module into the traditional network for domain shift compensation, as illustrated in Fig.~\ref{network_frame}(a). More details about the different components of our SGDA are given as follows.


\subsection{Slice Grouped SE Adapter}
We first begin by designing an extra light-weight adapter to compensate for domain shift. As demonstrated in \cite{wang2019towards}, the squeeze-and-excitation module \cite{8578843} can be seen as a feature-based attention mechanism for dealing with domain shift due to its channel-wise rescaling ability. Nonetheless, it is designed for 2D object detection, and directly applying it to pulmonary nodule detection in 3D CT scans may not be optimal \cite{9314699}. This is because that simply converting the 2D pooling operation of it to 3D will lead to severe loss of information. To solve this problem, we consider that, in the thoracic CT images, vessels and bronchus have the shape of continuous pipe, whereas nodules are usually isolated and spherical; thus in order to distinguish nodules from other tissues, doctors only need to view several consecutive slices to capture the relevance among them. Inspired by the diagnosis way of doctors, we propose a slice grouped squeeze-and-excitation (SGSE) adapter based on the squeeze-and-excitation module in \cite{8578843}. The SGSE mimics doctors to learn explicit channel interdependencies across consecutive slices in three different directions to modulate channel responses.

\begin{figure}[!t]
\flushleft
\setlength{\abovecaptionskip}{-0.01in}
\includegraphics[width=3in]{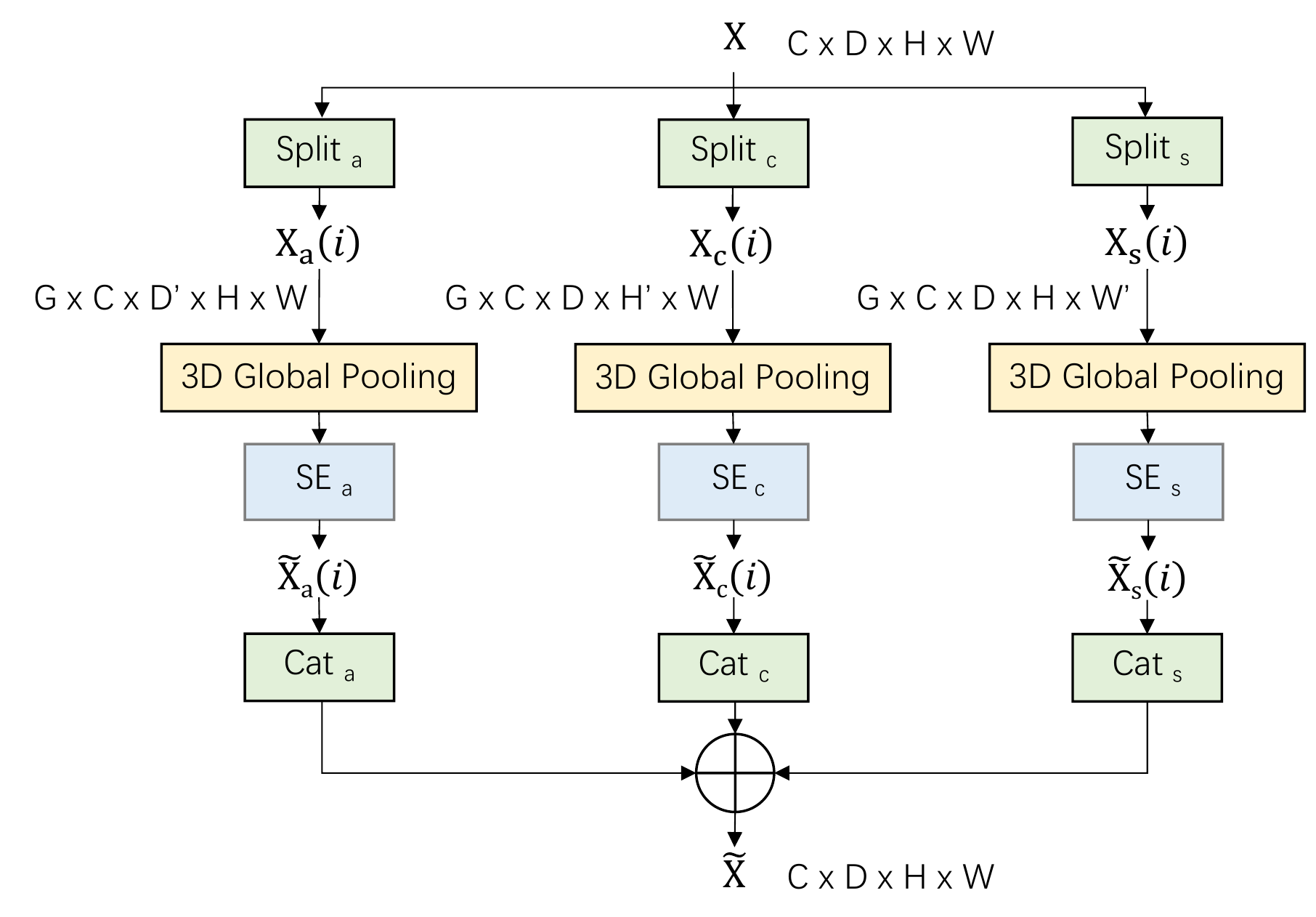}
\caption{Proposed Slice Grouped Squeeze-and-Excitation (SGSE) Adapter. It works in axial, coronal, and sagittal directions to project the input feature map along a subspace matched to the statistics of a particular domain by group, and then sums the three directional projections.}
\label{SGSE_adapter}
\end{figure}

As illustrated in Fig. \ref{SGSE_adapter}, the SGSE adapter consists of the sequences of operations across several adjacent slices in three directions, and the output feature is the mean of three directional output features. Let $ \mathbf{X} \in  \mathbb{R} ^ {C \times D \times H \times W}$ denote the input feature map for the SGSE adapter, where $D$, $H$, $W$, and $C$ represent depth, height, width, and the number of channels, respectively. We split the input feature map $\mathbf{X}$ into $G$ groups along the axial, coronal, and sagittal axis respectively, to obtain the slice grouped volumes in each direction: $ \mathbf{X}_a(i) \in  \mathbb{R} ^ {C \times D^\prime \times H \times W}$, $ \mathbf{X}_c(i) \in  \mathbb{R} ^ {C \times D \times H^\prime \times W}$, $ \mathbf{X}_s(i) \in  \mathbb{R} ^ {C \times D \times H \times W^\prime}$, where $i = 1,...,G$, $D^\prime = D / G$, $H^\prime = H / G$, and $W^\prime = W / G$. Each group is executed independently as following Eq.~(\ref{eq1}) to compute $\mathbf{Y}_a(i)$, $\mathbf{Y}_c(i)$, or $\mathbf{Y}_s(i)$ for channel response modulation, which is further applied to the group as Eq.~(\ref{eq2}) to compute $\widetilde{\mathbf{X}}_a(i)$, $\widetilde{\mathbf{X}}_c(i)$, or $\widetilde{\mathbf{X}}_s(i)$; then the results in the same direction are concatenated to obtain the directional output feature $\widetilde{\mathbf{X}}_a$, $\widetilde{\mathbf{X}}_c$, or $\widetilde{\mathbf{X}}_s$:
\begin{align}
\label{eq1}
\nonumber
\mathbf{Y}_{(\varphi)}(i) &= \mathbf{F}_{SE}(\mathbf{F}_{avg}(\mathbf{X}_{(\varphi)}(i)), \mathbf{W}_{(\varphi)1}, \mathbf{W}_{(\varphi)2}) \\ &= \mathbf{W}_{(\varphi)2}\delta(\mathbf{W}_{(\varphi)1}\mathbf{F}_{avg}(\mathbf{X}_{(\varphi)}(i))),
\\\label{eq2}
\widetilde{\mathbf{X}}_{(\varphi)}(i) &= \mathbf{F}_{scale}(\mathbf{X}_{(\varphi)}(i), \sigma(\mathbf{Y}_{(\varphi)}(i))),
\\
\label{eq3}
\widetilde{\mathbf{X}}_{(\varphi)} &= [\widetilde{\mathbf{X}}_{(\varphi)}(1),...,\widetilde{\mathbf{X}}_{(\varphi)}(G)] \in \mathbb{R}^{C \times D \times H \times W},
\end{align}
where $\mathbf{F}_{avg}$ is the 3D average pooling operation, and $\delta$ refers to the ReLU function. $\mathbf{W}_{(\varphi)1} \in \mathbb{R}^{\frac{C}{r} \times C}$ and $\mathbf{W}_{(\varphi)2} \in \mathbb{R}^{C \times \frac{C}{r}}$ are FC layers; $r$ denotes the channel dimension reduction factor. $\mathbf{W}_{(\varphi)1}$ and $\mathbf{W}_{(\varphi)2}$ are shared by different groups $\mathbf{X}_{(\varphi)}(i)$ in the same direction, and are distinct for different directions. $\varphi$ represents axial, coronal, or sagittal axis. $\sigma$ refers to the sigmoid function, and $\mathbf{F}_{scale}$ implements a channel-wise multiplication.
Finally, the output feature $\widetilde{\mathbf{X}}$ is
\begin{equation}
\widetilde{\mathbf{X}}_{SGSE} = (\widetilde{\mathbf{X}}_a + \widetilde{\mathbf{X}}_c + \widetilde{\mathbf{X}}_s) / 3.
\end{equation}

We note that the 3D version of squeeze-and-excitation module\cite{8578843} is a special case of our SGSE adapter, when we set the number of groups to be one.

\subsection{Slice Grouped Domain Attention}

Deep learning has advanced the state-of-the-arts for pulmonary nodule detection; however, existing detectors are usually customized and trained for a certain task/dataset. On the one hand, these detectors typically suffer from significant performance degradation on unseen datasets that have different distributions from the observed data. On the other hand, these detectors lack the capability of operating over multiple domains, which is actually a challenging topic in computer vision. We aim to design a universal pulmonary nodule detector robust for multiple datasets. To the best of our knowledge, this is the first time to learn a universal network for multi-center datasets in the field of pulmonary nodule detection.

In order to allow adaptation to different domains within a universal network, we introduce the abovementioned SGSE adapters as domain-specific layers, and the remainder of the network is shared across domains, as is commonly done in multi-domain learning. In addition, instead of using hard attention mechanism the same as multi-domain learning to force the network to fully attend to a single domain, inspired by \cite{wang2019towards}, we adopt the domain assignment mechanism of Fig. \ref{network_frame} (c). The overall flowchart of our proposed SGDA is exhibited in Fig. \ref{network_frame} (b), which consists of three branches of a universal SGSE adapter bank and its corresponding domain assignment mechanism. In this way, the domain can be inferred automatically from three directions, and more importantly, any of the tasks can be solved in any of the domains without prior knowledge of the tasks/datasets, since it is not necessary to limit domains according to the tasks/datasets. For example, the widely used nodule dataset LUNA16 \cite{DBLP:journals/mia/SetioTBBBCCDFGG17} is co-created by several academic centers and medical imaging companies, which can have many sub-domains, e.g. due to CT devices (GE Medical vs. Siemens), annotation habits of doctors, the severity of nodules (malignancy vs. benignity), radiation dose, etc. Another example is that, one sub-domain of LUNA16 \cite{DBLP:journals/mia/SetioTBBBCCDFGG17} may follow similar distribution as one sub-domain of tianchi \cite{tianchi}; thus these two sub-domains can be merged into one. Actually, the domains may not even have clear semantics, and they can be data-driven. Thus, the soft domain assignment mechanism proposed makes more sense.

\subsubsection{Universal SGSE Adapter Bank}
A universal pulmonary nodule detector should involve multi-domain information, thus being able to adapt to different domains. To realize this, we construct the universal SGSE adapter bank.

\begin{figure}[!t]
\centering
\setlength{\abovecaptionskip}{-0.01in}
\includegraphics[width=1.1in]{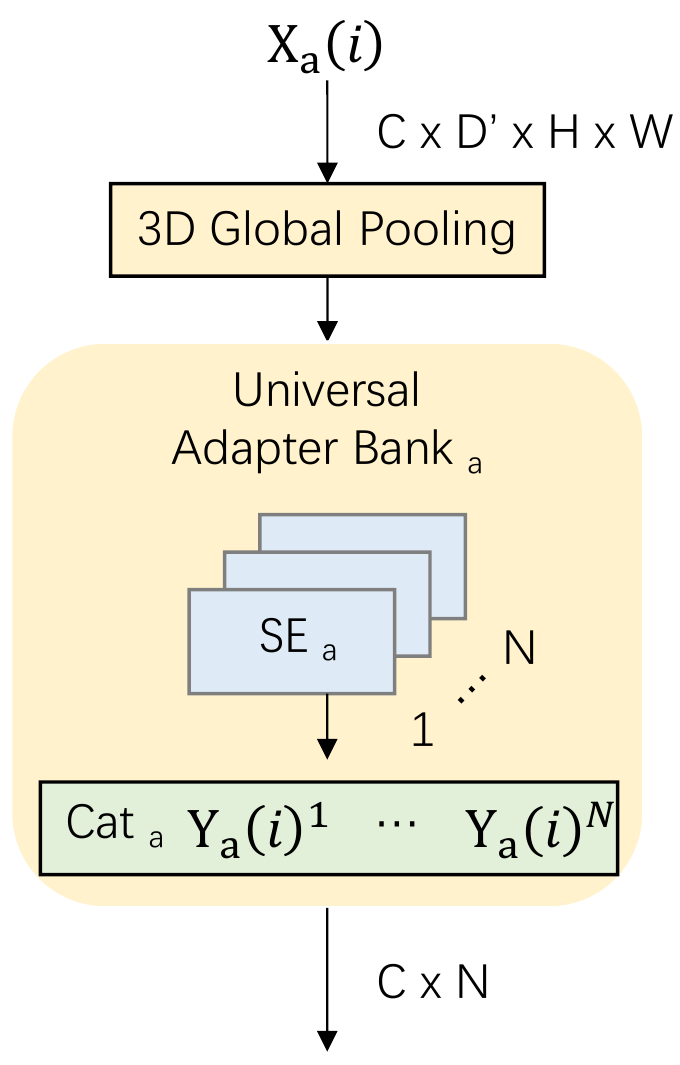}
\caption{Universal SGSE Adapter Bank in one direction. It projects the input group along $N$ subspaces of domains in this direction, and then concatenates these projections.}
\label{adapter_bank}
\end{figure}

The universal SGSE adapter bank of Fig. \ref{network_frame}(c) is a universal module composed of several single SGSE adapters, which are then integrated by direction as shown in Fig. \ref{adapter_bank}; each SGSE adapter corresponds to one domain. The universality of the universal SGSE adapter bank is implemented by concatenating one group's outputs of certain directional branch of the individual SGSE adapters to form a universal representation space for this group in the certain direction
\begin{equation}
\mathbf{Y}_{(\varphi)}(i)^{Uni} = [\mathbf{Y}_{(\varphi)}(i)^1,\mathbf{Y}_{(\varphi)}(i)^2,...,\mathbf{Y}_{(\varphi)}(i)^N] \in \mathbb{R}^{C \times N},
\end{equation}
where $N$ is the number of adapters, and $\mathbf{Y}_{(\varphi)}(i)^j$ is the output of the $j$-th adapter of the $i$-th group in the $\varphi$ direction, given by Eq.~(\ref{eq1}). Note that $N$ is a hyperparameter, and there is no need to make it identical to the number of tasks/datasets. Each SGSE adapter in the bank (non-linearly) projects the input by group along three directional data-driven subspaces of domains. The combination of these data-driven feature subspaces, which are later trained on all pulmonary nodule datasets, is thus able to cover the feature subspaces of the domains spanned by all the datasets. Then the concatenation of the $N$ projections in certain direction of a group is fed in the corresponding directional domain assignment component to reweight each of these projections for combination also in a data-driven way.

\subsubsection{Domain Assignment}
The domain assignment component shown in Fig. \ref{network_frame}(c) correspondingly also works in three directions to produce a domain-sensitive set of weights for each group to combine the directional SGSE adapter projections. Following \cite{wang2019towards}, each directional domain assignment component first applies a 3D global average pooling to the $i$-th input group in this direction to remove spatial dimensions, and then a softmax layer (linear layer plus softmax function)
\begin{align}
\nonumber
\mathbf{Y}_{(\varphi)}(i)^{DA} &= \mathbf{F}_{DA}(\mathbf{X}_{(\varphi)}(i)) \\
&= softmax(\mathbf{W}_{(\varphi),DA} \mathbf{F}_{avg}(\mathbf{X}_{(\varphi)}(i))),
\end{align}
where $\mathbf{W}_{(\varphi),DA} \in \mathbb{R}^{N \times C}$ is FC layer. It varies for different directions, while keeps the same for groups in the same direction. Then the vector $\mathbf{Y}_{(\varphi)}(i)^{DA}$ is used to combine the outputs of the corresponding directional universal SGSE adapter bank after the $i$-th input group entered, to obtain a vector of domain adaptive responses for this group
\begin{equation}
\mathbf{Y}_{(\varphi)}(i) = \mathbf{Y}_{(\varphi)}(i)^{Uni}\mathbf{Y}_{(\varphi)}(i)^{DA}.
\end{equation}
Subsequently, as Eq.~(\ref{eq2}) in the SGSE adapter, $\mathbf{Y}_{(\varphi)}(i)$ is used to modulate channel responses of the $i$-th input group in this direction, e.g. the $\varphi$ direction. Finally, the modulated groups in the same direction are concatenated as Eq.~(\ref{eq3}) to obtain the directional output feature map $\widetilde{\mathbf{X}}_a$, $\widetilde{\mathbf{X}}_c$, or $\widetilde{\mathbf{X}}_s$. Compared with the outputs of the SGSE adapter, the directional output feature maps here involve a wealth of multi-domain information; simple summation of them seems to be a bad choice. Therefore, we propose a three way cross attention to fuse the output feature maps in three directions.

\begin{figure}[!t]
\centering
\setlength{\abovecaptionskip}{-0.01in}
\includegraphics[width=2.5in]{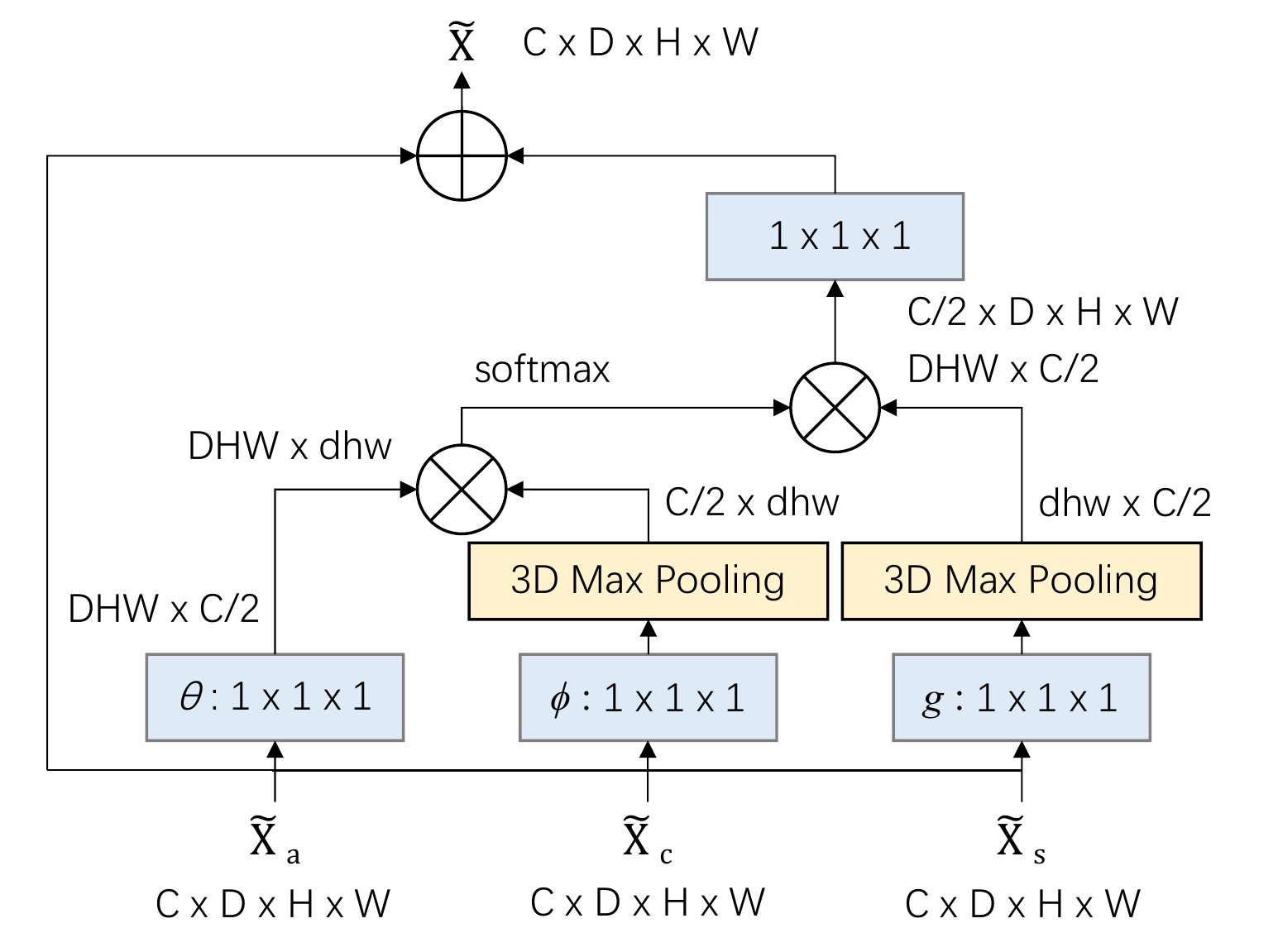}
\caption{Three Way Cross Attention. Feature maps in two of the three directions are computed to obtain the weights on the third one. The shapes of the feature maps after certain operations are shown.}
\label{cross_attention}
\end{figure}

\subsubsection{Three Way Cross Attention}

As mentioned above, rather than merely summing the directional modulated feature maps, we further design a three way cross attention module as shown in Fig. \ref{cross_attention} for their fusion. We first compute the similarity of two feature maps $\widetilde{\mathbf{X}}_a$ and $\widetilde{\mathbf{X}}_c$ using embedded dot product, and then apply a softmax function to obtain the weights on the other feature map $\widetilde{\mathbf{X}}_s$ after it gets embedded. Afterwards, the output is embedded for addition to the summation of $\widetilde{\mathbf{X}}_a$, $\widetilde{\mathbf{X}}_c$, and $\widetilde{\mathbf{X}}_s$:
\begin{align}
\label{eq8}
\widetilde{\mathbf{X}} & = (\widetilde{\mathbf{X}}_a + \widetilde{\mathbf{X}}_c + \widetilde{\mathbf{X}}_s) / 3 + \mathbf{W}_{CA}\mathbf{Y}_{CA}, \\
\label{eq9}
\mathbf{Y}_{CA} &= softmax(\widetilde{\mathbf{X}}_a^T\mathbf{W}_\theta^T\mathbf{W}_\phi\widetilde{\mathbf{X}}_c)\widetilde{\mathbf{X}}_s^T\mathbf{W}_g^T,
\end{align}
where $\mathbf{W}_{CA}$, $\mathbf{W}_\theta$, $\mathbf{W}_\phi$, and $\mathbf{W}_g$ are $1\times1\times1$ convolutional layers. In order to reduce the computation, the number of channels represented by $\mathbf{W}_\theta$, $\mathbf{W}_\phi$, and $\mathbf{W}_g$ are set to be half of those in $\widetilde{\mathbf{X}}_{(\varphi)}$. $\mathbf{W}_{CA}$ later restores the number of channels for matrix addition. We also add max pooling layers after $\mathbf{W}_\phi$ and $\mathbf{W}_g$ as Fig. \ref{cross_attention}.

\begin{table*}[!t]
\renewcommand{\arraystretch}{1}
\caption{Pulmonary nodule datasets. `Scans' denotes the number of CT scans. `Nodules' denotes the number of labeled nodules. `Class' denotes the class number. And `Raw' means whether the dataset contains raw CT scans. `Image Size' gives the dimensions of the CT image matrix along the x, y, and z axes. `Spacing' gives the voxel sizes (mm) along the x, y, and z axes.}
\label{dataset_info}
\centering
\setlength{\tabcolsep}{1.4mm}{
\begin{tabular}{l | c c c c c l l l} \hline
    Dataset & Year & Scans & Nodules & Class & Raw & File Size & Image Size & Spacing\\ \hline
    LUNA16 \cite{DBLP:journals/mia/SetioTBBBCCDFGG17} & 2016 & 601 & 1186 & 2 & Yes & 25M-258M & $512\times512\times95-512\times512\times733$ & $(0.86,0.86,2.50)-(0.64,0.64,0.50)$ \\
    tianchi \cite{tianchi} & 2017 & 800 & 1244 & 2 & Yes & 26M-343M & $512\times512\times114-512\times512\times1034$ & $(0.66,0.66,2.50)-(0.69,0.69,0.30)$ \\
    russia \cite{russia} & 2018 & 364 & 1850 & 2 & Yes & 80M-491M & $512\times512\times313-512\times512\times1636$ & $(0.62,0.62,0.80)-(0.78,0.78,0.40)$ \\
    PN9 \cite{mei2021sanet} & 2021 & 8796 & 40436 & 9 & No & 5.6M-73M & $212\times212\times181-455\times455\times744$ & $(1.00,1.00,1.00)-(1.00,1.00,1.00)$ \\ \hline
    \end{tabular}}
\end{table*}

Due to the large computational cost of the pulmonary nodule detection task, although we already use some tricks above, our proposed three way cross attention module may still not be able to fit in few 3D pulmonary nodule detection networks. Therefore, a grouping trick as in \cite{Xception17}, \cite{mobilenets17}, \cite{ResNeXt17}, \cite{GN18}, \cite{crossformer21}, can be used to further reduce computation. We divide the three embedded directional feature maps (after pooling) all along the depth dimension into $G$ groups, each of which contains $D^\prime = D / G$ or $d^\prime = d / G$ depths of its corresponding feature map, and then employ the three way cross attention in every three matching groups. The Eq.~(\ref{eq9}) is modified as
\begin{equation}
\mathbf{Y}_{CA}(i) = softmax((\widetilde{\mathbf{X}}_a^T\mathbf{W}_\theta^T)(i)(\mathbf{W}_\phi\widetilde{\mathbf{X}}_c)(i))(\widetilde{\mathbf{X}}_s^T\mathbf{W}_g^T)(i).
\end{equation}
The outputs of these groups are concatenated in the axial direction/along the depth dimension for addition to the summation of the three original directional feature maps.

\begin{table}[!t]
\renewcommand{\arraystretch}{1}
\caption{Pulmonary nodule size distribution of datasets. `$d$' denotes the diameter (mm).}
\label{nodule_size}
\centering
\setlength{\tabcolsep}{0.6mm}{
\begin{tabular}{l | c c c c c | c} \hline
    Dataset & $d < 3$ & $3 \leq d < 5$ & $5 \leq d < 10$ & $10 \leq d < 30$ & $30 \leq d$ & All\\ \hline
    LUNA16 \cite{DBLP:journals/mia/SetioTBBBCCDFGG17} & - & 270 & 635 & 279 & 2 & 1186 \\
    tianchi \cite{tianchi} & 1 & 213 & 596 & 423 & 11 & 1244 \\
    russia \cite{russia} & 6 & 552 & 907 & 360 & 25 & 1850 \\
    PN9 \cite{mei2021sanet} & 9 & 4678 & 29213 & 6053 & 483 & 40436 \\ \hline
    \end{tabular}}
\end{table}

All in all, the feature subspaces of the domains spanned by all pulmonary nodule datasets are captured from the axial, coronal, and sagittal directions by the universal SGSE adapter bank of the domain attention module; then the domain assignment component soft-routes the combination of the output projections of the bank by group in each direction. Both operations are data-driven, and do not require prior knowledge of the domain. In this way, our proposed SGDA achieves universal pulmonary nodule detection over multiple datasets. What's more, implementing this module allows the network to leverage shared knowledge across domains, which further improves the performance of the network. Note that the output layer has to be task/dataset-specific, since different pulmonary nodule datasets may use different annotation standards.

\subsection{Discussion}
It is noteworthy that the 3D version of \cite{wang2019towards} is a special case of our approach, when the number of groups in the SGDA module is 1. However, our approach substantially differs from \cite{wang2019towards} in the following aspects: (1) Their work focuses on universal object detection using 2D CNN networks, while we aim to realize 3D universal pulmonary nodule detection in the pure medical field. The two are fundamentally different in both network structures and tasks. (2) We propose a novel SGSE adapter/form a new SGSE adapter bank according to the characteristics of nodule detection in CT images, and also a three way cross attention to fuse the output feature maps in three directions of our SGDA. (3) They mainly explore their method's capability of working on multiple domains. Nonetheless, we not only study this, but also invest significant efforts in validating the generalization ability of the nodule detection networks after our proposed SGDA module is plugged.


\section{Experiments}
\label{sec:Experiments}

In this section, we conduct extensive experiments to investigate the effectiveness of our proposed SGDA module on multiple pulmonary nodule datasets.

\subsection{Datasets and Evaluation}
Our experiments are mainly conducted on four pulmonary nodule datasets: LUNA16 \cite{DBLP:journals/mia/SetioTBBBCCDFGG17}, tianchi \cite{tianchi}, russia \cite{russia}, and PN9 \cite{mei2021sanet}. The details of these datasets are shown in the Table~\ref{dataset_info}. We also present the pulmonary nodule size distribution of the four datasets in Table~\ref{nodule_size}. The CT scans in these datasets are collected from diverse sites with various thicknesses. We consider only those CT scans with annotations of nodule locations. The annotation files of LUNA16, tianchi, and russia are csv files containing one nodule per line. Each line holds the filename of the CT scan, the center coordinates, and the diameter of one nodule. PN9 has a similar annotation file, except for using top-left and bottom-right coordinates to denote nodule location.

We use the Free-Response Receiver Operating Characteristic (FROC), which is the official evaluation metric of the widely used pulmonary nodule dataset LUNA16, for evaluation in all cases. It is defined as the average recall rate at 0.125, 0.25, 0.5, 1, 2, 4, and 8 false positives per scan. A nodule candidate is regarded as a true positive if it is located within a distance $R$ from the center of any annotated nodules, where $R$ denotes the radius of the annotated nodule. Nodule candidates not located in the range of any annotated nodules are regarded as false positives. All the nodule candidates are evaluated on their corresponding dataset respectively. Through this, all detectors are evaluated on their corresponding dataset, because different tasks/datasets use different output layers. All in all, we evaluate the detector for each task/dataset via the aforementioned FROC, and the mean value of the FROCs is used to measure the universal pulmonary nodule detection performance. 

\begin{figure}[!t]
\centering
\setlength{\abovecaptionskip}{-0.001in}
\includegraphics[width=2.6in]{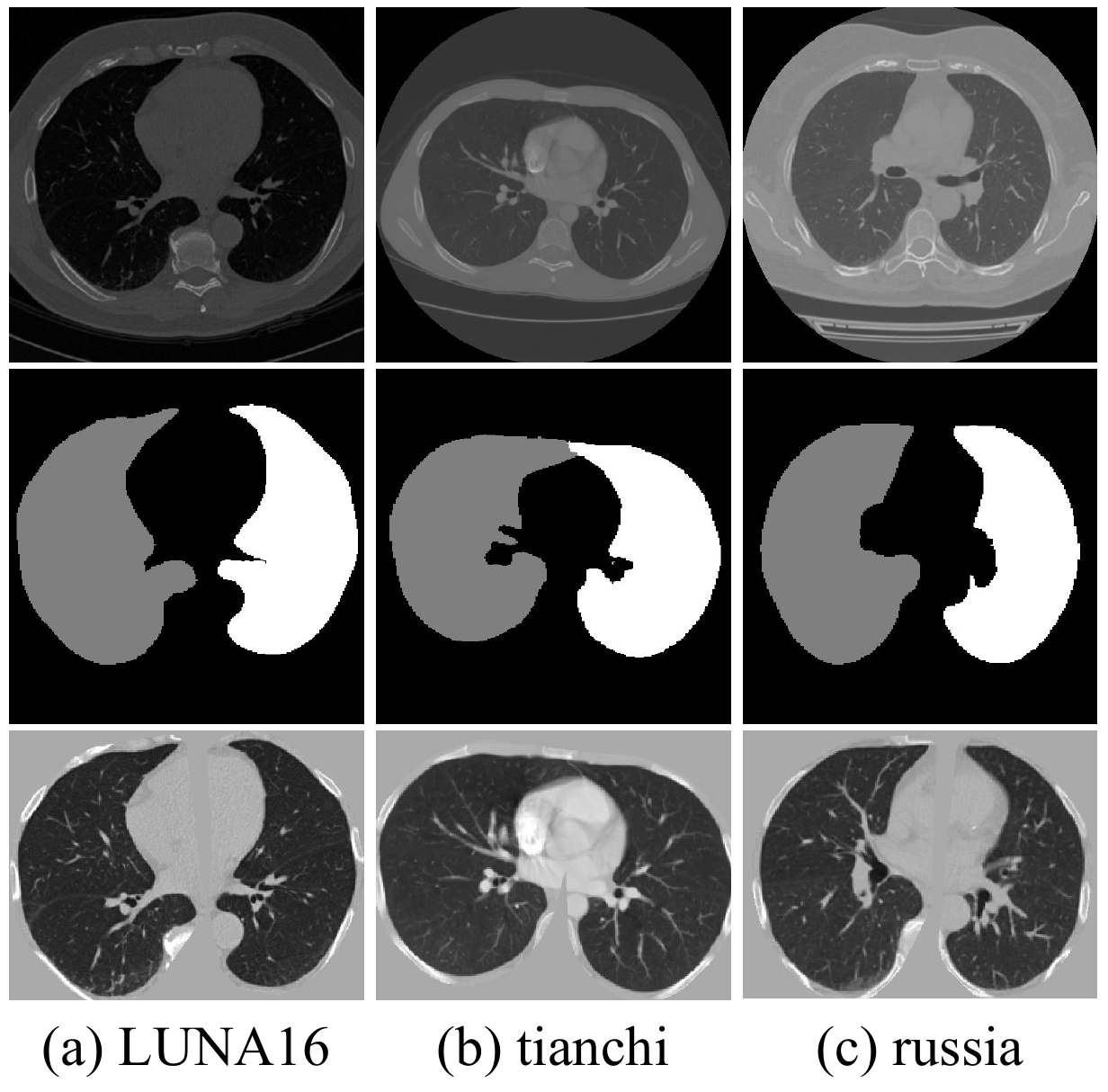}
\caption{Samples of the preprocessed images in the LUNA16, tianchi, and russia. The 1st row are the raw images, the 2nd row are the segmented lung regions, and the 3rd row are the preprocessed images, respectively.}
\label{preprocess}
\end{figure}


\subsection{Data Preprocessing}

\begin{table*}[!t]
\setlength{\abovecaptionskip}{-0.001in}
\renewcommand{\arraystretch}{1}
\caption{Comparison of our SGDA and other multi-domain methods in terms of FROC on dataset LUNA16, tianchi, and russia. Values below the names of datasets are FROCs (unit: \%). All the methods utilize NoduleNet as backbone: (1) shared models with the prefix `uni-', (2) independent models with the word `single' in the name, (3) multi-domain methods, (4) universal models with `SG' in the name (Ours).}
\label{experimental_results}
\centering
\setlength{\tabcolsep}{3.7mm}
\begin{minipage}{18cm}
\begin{tabular}{l|c|c|c|c|c c c|c} \hline
    Method & \#Adapters & \#Groups & \#Params & \#FLOPs\footnote{The number of FLOPs of the backbone networks is counted using the fvcore toolkit \cite{flops_cal}. The computation complexity of the detection heads is not counted since they are the same among different methods.} & LUNA16 & tianchi & russia & Avg\\ \hline
    single NoduleNet \cite{10.1007/978-3-030-32226-7_30} & - & - & 16.73M$\times$3 & 139G & 77.71 & 68.23 & 37.19 & 61.04 \\
    uniNoduleNet & - & - & 39.50M & 139G & 79.88 & 68.60 & 33.35 & 60.61 \\ \hline
    NoduleNet+BN \cite{DBLP:journals/corr/BilenV17} & 3 & - & 39.51M & 139G & 79.94 & 68.12 & 36.52 & 61.52 \\
    NoduleNet+series \cite{DBLP:conf/cvpr/RebuffiBV18} & 3 & - & 40.14M & 145G & 78.44 & 70.41 & 33.39 & 60.74 \\
    NoduleNet+parallel \cite{DBLP:conf/cvpr/RebuffiBV18} & 3 & - & 40.13M & 145G & 78.57 & 70.14 & 35.61 & 61.44 \\
    NoduleNet+separable \cite{DBLP:conf/miccai/HuangHYZZ19} & 3 & - & 34.68M & 13.58G & 66.31 & 62.26 & 32.96 & 53.84 \\
    NoduleNet+SNR \cite{DBLP:journals/corr/abs-2101-00588} & - & - & 39.50M & 139G & 69.52 & 66.57 & 36.76 & 57.61 \\ \hline
    single NoduleNet+SE \cite{wang2019towards} & - & - & 16.74M$\times$3 & 139G & 77.78 & 68.86 & 38.06 & 61.56 \\
    uniSENoduleNet \cite{wang2019towards} & - & - & 39.51M & 139G & 80.53 & 69.13 & 34.34 & 61.33 \\
    NoduleNet+SE \cite{wang2019towards} & 3 & - & 39.54M & 139G & 78.89 & 72.33 & 35.89 & 62.37 \\
    DANoduleNet \cite{wang2019towards} & 3 & - & 39.54M & 139G & \textbf{82.63} & 73.29 & 38.50 & 64.80 \\ \hline
    single NoduleNet+SGSE & - & 4 & 16.77M$\times$3 & 139G & 78.30 & 70.36 & \textbf{39.01} & 62.55 \\
    uniSGSENoduleNet & - & 4 & 39.54M & 139G & 81.12 & 71.00 & 38.42 & 63.51 \\
    NoduleNet+SGSE & 3 & 4 & 39.62M & 139G & 80.93 & 70.94 & 38.30 & 63.39 \\
    SGDANoduleNet & 3 & 4 & 39.82M & 147G & 81.91 & \textbf{77.13} & 37.15 & \textbf{65.39} \\ \hline
    \end{tabular}
    \end{minipage}
\end{table*}

LUNA16 \cite{DBLP:journals/mia/SetioTBBBCCDFGG17}, tianchi \cite{tianchi}, and russia \cite{russia} are split into 7/1/2 for training, validation, and testing. There are three preprocessing steps for the raw CT data in these three datasets. First, in order to reduce the irrelevant calculation, we segment lung regions from each CT image using lungmask \cite{DBLP:journals/corr/abs-2001-11767}, and after converting the raw data from Hounsfield Unit (HU) to unit8, we assign the regions other than the lung masks a padding value 170. Thereinto, The HU values are clipped into $[-1200,600]$, and transformed linearly into $[0,255]$ to obtain unit8 values. Second, to avoid too much unnecessary hyperparameters, we resample all the CT images into $1\times1\times1$ mm spacing to keep anchors in all the detectors consistent. Third, in order to further reduce the computation, we utilize the segmented lung masks to restrict the size of the CT images. The preprocessed CT images are as shown in Fig.~\ref{preprocess}. For PN9 \cite{mei2021sanet} dataset, we use the same data preprocessing as \cite{mei2021sanet}. We adopt voxel coordinates for all the cases, and modify the annotation coordinates according to our preprocessing processes.

During training, it is not feasible for 3D CNN to use the entire CT images as input due to the limitation of GPU memory. Thus, small 3D patches are extracted from the CT images and individually input into the network. The size of the extracted 3D patch is $1\times128\times128\times128$ (Channel$\times$Depth$\times$Height$\times$Width). If a patch exceeds the range of the CT image from which it is extracted, it is padded with a value of 170. During testing, the entire CT images are input into the network without being cropped into 3D patches. In order to avoid an odd size of the entire CT images, they are padded with a value of 170 before being input into the network.

\subsection{Small-Scale Experiments}

In our small-scale experiments, we utilize the widely used pulmonary nodule detection network NoduleNet \cite{10.1007/978-3-030-32226-7_30} as the backbone. The small-scale experiments are done on the LUNA16 \cite{DBLP:journals/mia/SetioTBBBCCDFGG17}, tianchi \cite{tianchi}, and russia \cite{russia} datasets, which are all split into 7/1/2 for training, validation, and testing. The multi-domain methods are trained from scratch on all datasets of interest, e.g. LUNA16 \cite{DBLP:journals/mia/SetioTBBBCCDFGG17}, tianchi \cite{tianchi}, and russia \cite{russia}, simultaneously. All inputs of a batch are from a single (randomly sampled) pulmonary nodule dataset, and in each epoch, all 3D patches of each dataset are processed only once. We use the Stochastic Gradient Descent (SGD) optimizer with a batch size of 8. The initialization learning rate is set to 0.01; the momentum and weight decay coefficients are set to 0.9 and $1\times10^{-4}$, respectively. The learning rate decreases to 0.001 after 200 epochs and 0.0001 after another 120 epochs. The NoduleNet has many hyperparameters, and we use the same hyperparameters as NoduleNet\footnote{\url{https://github.com/uci-cbcl/NoduleNet/}} across datasets for all networks except for the number of epochs of training without RCNN. It is tuned for each network to obtain the best performance. All the small-scale experiments are implemented using PyTorch on 1 NVIDIA GeForce RTX 3090 GPU with 24GB memory.

\begin{figure}[!t]
\centering
\setlength{\abovecaptionskip}{-0.01in}
\includegraphics[width=3.5in]{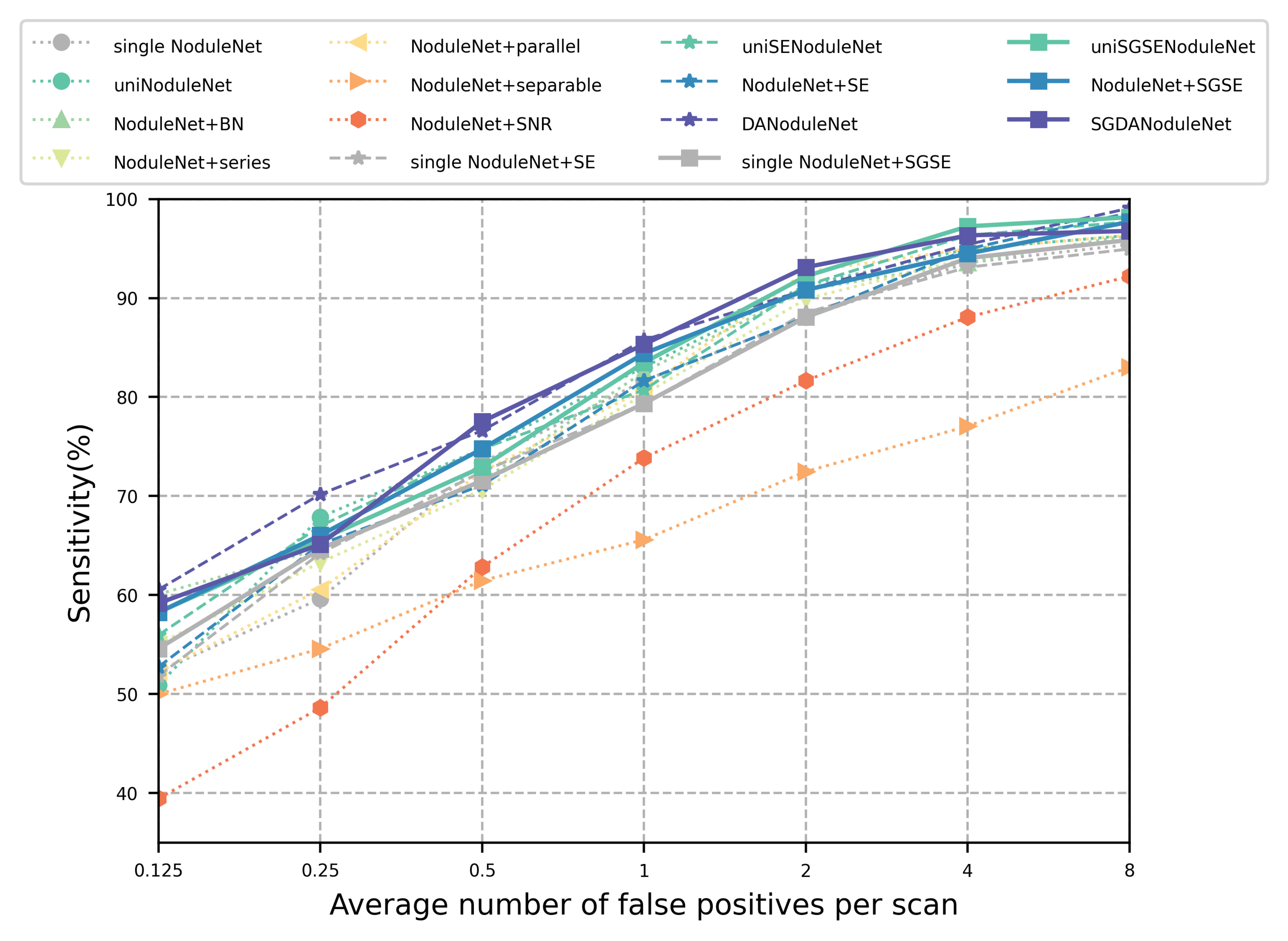}
\caption{FROC curves of compared methods and our SGDA on LUNA16.}
\label{luna16_FROC}
\end{figure}

\subsubsection{Multi-center Experiments}
\label{small_multi}

Here we perform experiments to evaluate the proposed SGDA in dealing with multi-center pulmonary nodule detection using LUNA16 \cite{DBLP:journals/mia/SetioTBBBCCDFGG17}, tianchi \cite{tianchi}, and russia \cite{russia}. Specifically, we compare with the following approaches:
\begin{itemize}
\item Models that have `single' in the name: train a model for each dataset separately;
\item Models that have the prefix `uni-': union different datasets into one dataset for training;
\item The multi-domain methods\footnote{We follow the code from \url{https://github.com/microsoft/SNR} for SNR implementation.};
\item Our universal models that have `SG' in the names.
\end{itemize}
In addition, to better demonstrate the superiority of our proposed method, we also conduct experiments on the first two set of approaches adding with a single SE or SGSE adapter (models that have `+SE' or `+SGSE' suffixes).


\begin{figure*}[!t]
\centering
\begin{minipage}[!t]{0.48\textwidth}
\centering
\setlength{\abovecaptionskip}{-0.01in}
\includegraphics[width=3.5in]{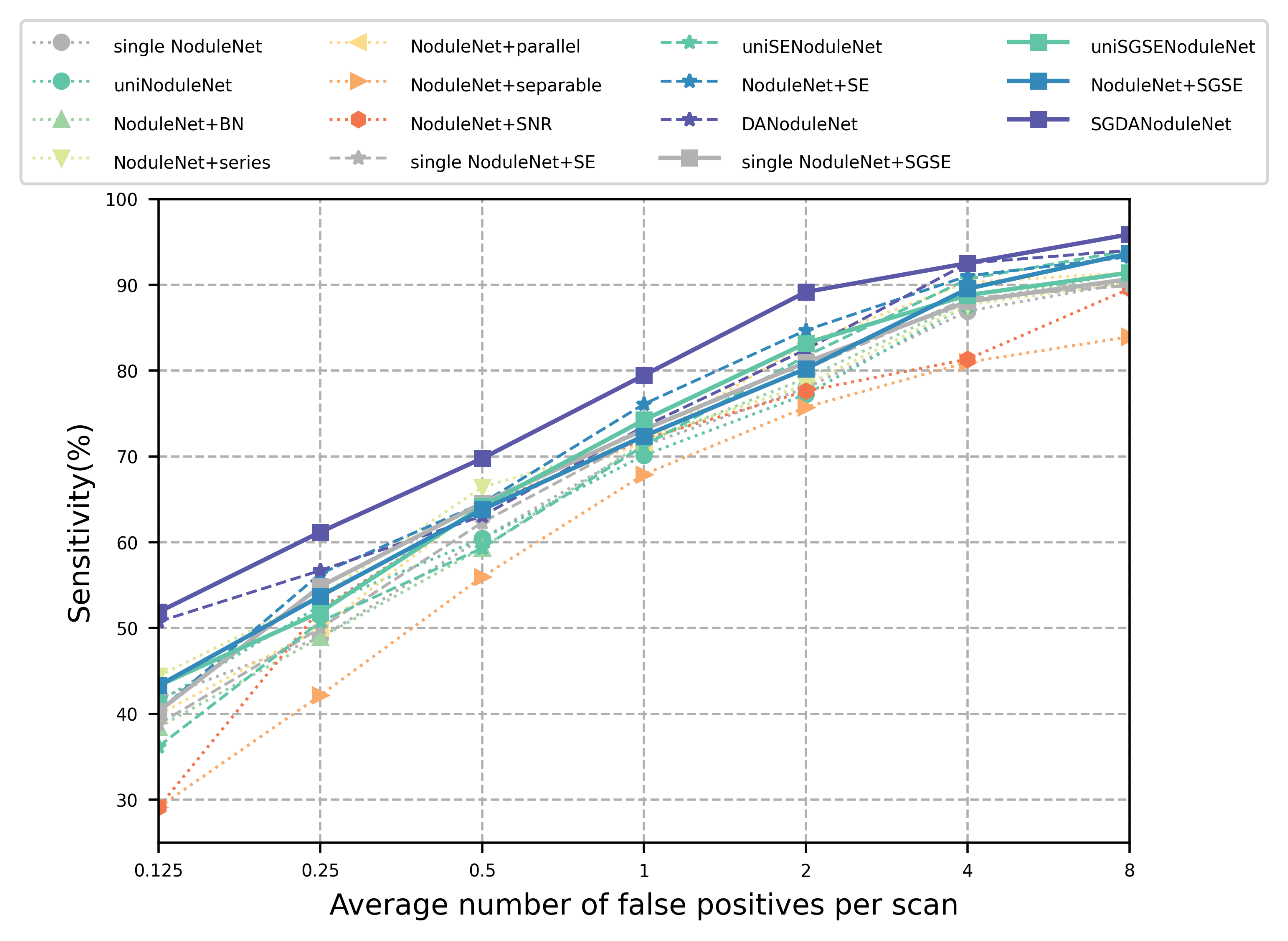}
\caption{FROC curves of compared methods and our SGDA on tianchi.}
\label{tianchi_FROC}
\end{minipage}
\begin{minipage}[!t]{0.48\textwidth}
\centering
\setlength{\abovecaptionskip}{-0.01in}
\includegraphics[width=3.5in]{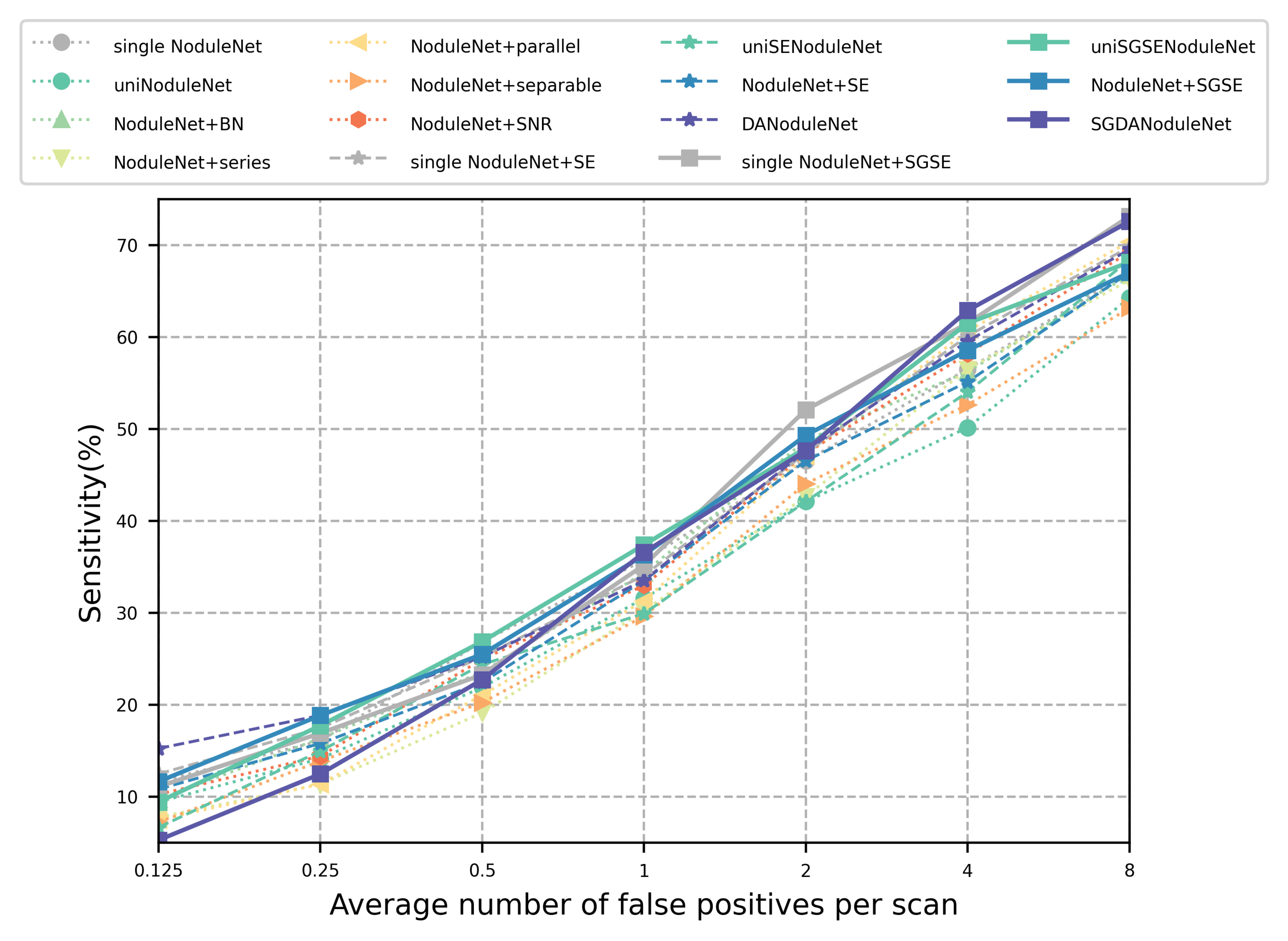}
\caption{FROC curves of compared methods and our SGDA on russia.}
\label{russia_FROC}
\end{minipage}
\end{figure*}




\begin{figure*}[!t]
\centering
\setlength{\abovecaptionskip}{-0.01in}
\includegraphics[width=7in]{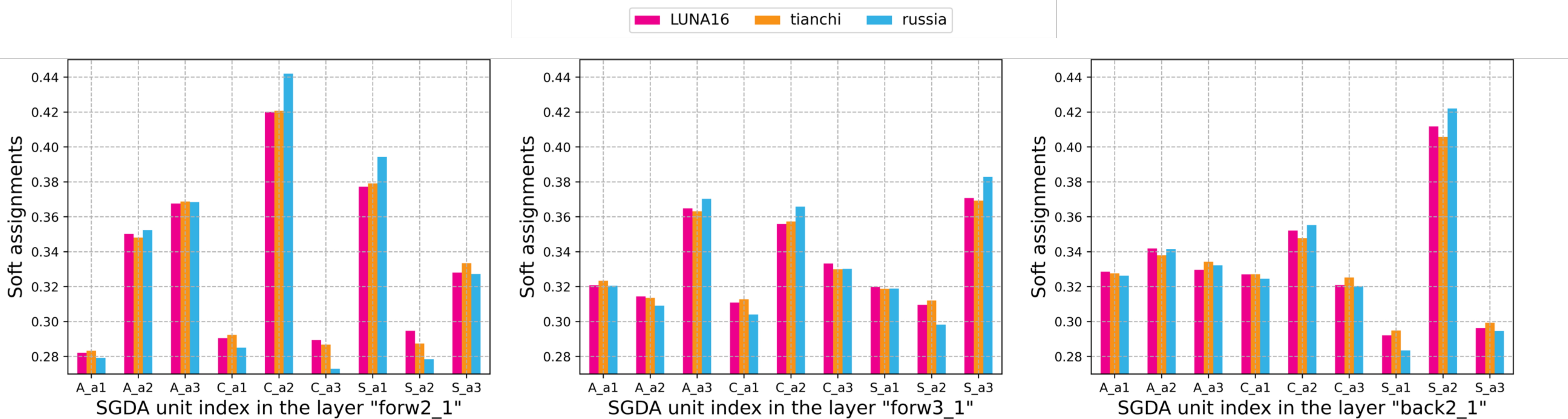}
\caption{Soft assignments on the SGSE adapters in axial, coronal, and sagittal directions inside our SGDA for the small-scale multi-center datasets, where ``forw2\_1'', ``back2\_1'', and ``forw3\_1'' denote the first residual blocks of the second forward stage and the corresponding backward stage, as well as the third forward stage, respectively.}
\label{visfeature_2_hist}
\end{figure*}

The overall experimental results are reported in Table~\ref{experimental_results}.
The detailed FROC curves for the three datasets are shown in Fig. \ref{luna16_FROC}, Fig. \ref{tianchi_FROC}, and Fig. \ref{russia_FROC} respectively. It is noted that our SGDA achieves the highest performance in most cases, and obtains an improvement of approximately 5\% over the simple ``uniNoduleNet'' baseline.
All the 2D models in \cite{wang2019towards} are re-implemented to be 3D for fair comparison, and these 3D versions are actually special cases of our models.


\textbf{Analysis.} To investigate how our proposed SGDA works in practice, we report the soft domain assignments on the SGSE adapters by direction inside each SGDA module for each dataset; the soft assignments are averaged over all test samples for each dataset. Some results are shown in Fig. \ref{visfeature_2_hist}. It is observed that different datasets, LUNA16 \cite{DBLP:journals/mia/SetioTBBBCCDFGG17}, tianchi \cite{tianchi}, and russia \cite{russia}, have different assignment distributions, and the russia¡¯s \cite{russia} assignments deviate much from those of the other two datasets. This observation reveals that there exist unneglectable domain shifts between the multi-center pulmonary nodule datasets, and our SGDA is able to give adaptive domain as-signments for different datasets. Currently, our work does not consider pushing the domain shifts in the adapters further apart and pulling together the remaining domain overlap, which may further enhance the generalization ability of the pulmonary nodule detection networks, and could be our future work.

\begin{table}[!t]
\setlength{\abovecaptionskip}{-0.001in}
\renewcommand{\arraystretch}{1}
\caption{Ablation study for the proposed SGDA module with different direction (\%).}
\label{different_direction}
\centering
\setlength{\tabcolsep}{2.2mm}{
\begin{tabular}{c|c|c c c|c} \hline
    Dir & \#Params & LUNA16 & tianchi & russia & Avg\\ \hline
    A & 39.54M & 79.61 & 69.77 & 37.00 & 62.12 \\  
    C & 39.54M & 80.66 & 68.76 & 35.61 & 61.67 \\ 
    S & 39.54M & 81.19 & 71.16 & 36.16 & 62.83 \\ 
    ACS w/o CA & 39.63M & 81.12 & 69.98 & \textbf{41.11} & 64.07 \\
    ACS & 39.82M & \textbf{81.91} & \textbf{77.13} & 37.15 & \textbf{65.39} \\ \hline
    \end{tabular}}
\end{table}


\begin{table}[!t]
\setlength{\abovecaptionskip}{-0.001in}
\renewcommand{\arraystretch}{1}
\caption{Ablation study for the proposed SGDA module with different number of groups $G$ (\%).}
\label{num_groups}
\centering
\setlength{\tabcolsep}{2.8mm}{
\begin{tabular}{c|c|c c c|c} \hline
    \#Groups & \#Params & LUNA16 & tianchi & russia & Avg\\ \hline
    1 & 39.54M & \textbf{82.63} & 73.29 & \textbf{38.50} & 64.80 \\ 
    4 & 39.82M & 81.91 & \textbf{77.13} & 37.15 & \textbf{65.39} \\ %
    8 & 39.82M & 77.78 & 72.06 & 36.56 & 62.13 \\ \hline
    \end{tabular}}
\end{table}

\begin{table}[!t]
\setlength{\abovecaptionskip}{-0.001in}
\renewcommand{\arraystretch}{1}
\caption{Ablation study for the proposed SGDA module with different number of adapters $N$ (\%).}
\label{num_adapters}
\centering
\setlength{\tabcolsep}{2.5mm}{
\begin{tabular}{c|c|c c c|c} \hline
    \#Adapters & \#Params & LUNA16 & tianchi & russia & Avg\\ \hline
    1 & 39.54M & 81.12 & 71.00 & \textbf{38.42} & 63.51 \\ 
    3 & 39.82M & \textbf{81.91} & \textbf{77.13} & 37.15 & \textbf{65.39} \\ %
    5 & 39.91M & 80.79 & 73.66 & 37.87 & 64.10 \\ \hline
    \end{tabular}}
\end{table}

\begin{table}[!t]
\setlength{\abovecaptionskip}{-0.001in}
\renewcommand{\arraystretch}{1}
\caption{Ablation study for the proposed SGDA module with different location of adapters (\%).}
\label{loc_adapters}
\centering
\setlength{\tabcolsep}{2.8mm}{
\begin{tabular}{c|c|c c c|c} \hline
    Location & \#Params & LUNA16 & tianchi & russia & Avg\\ \hline
    bottom & 39.53M & 79.48 & 75.05 & 37.63 & 64.05 \\
    middle & 39.58M & 79.22 & 70.57 & \textbf{39.05} & 62.94 \\
    top & 39.70M & 80.20 & 71.48 & 38.10 & 63.26 \\
    all & 39.82M & \textbf{81.91} & \textbf{77.13} & 37.15 & \textbf{65.39} \\ \hline %
    \end{tabular}}
\end{table}

\textbf{Different Direction.} To verify the necessity of performing split in three directions, we split the input feature map into groups only in one direction, i.e., axial, coronal, or sagittal respectively. Table~\ref{different_direction} shows the results of each of these three configurations and our proposed SGDA. The result of SGDA without cross attention is also involved for comparison. These results validate that both the grouping in different directions and cross attention are helpful for multi-center pulmonary nodule detection.

\textbf{Number of Groups.} We analyze the impact of different numbers of groups $G$ in our SGDA module, and the results are presented in Table~\ref{num_groups}. For a single group, since cross-attention is not necessary, the SGDA module reduces to a 3D DA module in \cite{wang2019towards}. From the results, we observe that the best performance is achieved when adopting 4 groups. These experimental results indicate that a proper number of groups improves the performance, but not the larger the better. Too many groups may degrade the capture of feature subspaces of the domains, and limit the detection ability of large nodules.

\textbf{Number of SGSE Adapters.} For the universal bank of Fig.~\ref{adapter_bank}, the number $N$ of SGSE adapters does not have to match the number of pulmonary nodule datasets. Table~\ref{num_adapters} summarises how the performance of the SGDA module are affected by $N$. For a single adapter, the SGDA module reduces to a SGSE adapter, and the universal detector reduces to a shared network. The results indicate that employing 3 adapters is preferable.

\textbf{Location of SGDA.} We investigate the cases of adding the SGDA module to different residual blocks/locations: to the first 4 residual blocks of the backbone encoder (bottom), to the last 6 residual blocks of the backbone encoder (middle), and to the 6 residual blocks of the backbone decoder (top), and to all the NoduleNet backbone residual blocks. Table~\ref{loc_adapters} shows that when SGDA is added to each residual block in the NoduleNet backbone, we achieve the best performance.

\begin{table}[!t]
\setlength{\abovecaptionskip}{-0.001in}
\renewcommand{\arraystretch}{1}
\caption{FROC (\%) of the compared methods and our SGDA finetuned on target dataset LUNA16 w.r.t percentage of training images. The first column is the result of a single NoduleNet directly trained on the whole target training set. The other values are FROCs with each column representing the percentage of training images.}
\label{cross1_luna16_table}
\centering
\setlength{\tabcolsep}{1.3mm}{
\begin{tabular}{l|c|c c c c c} \hline
    Method & LUNA16 & 20\% & 40\% & 60\% & 80\% & 100\% \\ \hline
    uniNoduleNet & 77.71 & 78.11 & 79.16 & 80.40 & 80.20 & 80.73 \\
    DANoduleNet \cite{wang2019towards} & - & 78.17 & 79.68 & 79.61 & 80.14 & \textbf{80.99} \\
    SGDANoduleNet & - & \textbf{80.01} & \textbf{80.27} & \textbf{80.79} & \textbf{80.66} & 80.20 \\ \hline
    \end{tabular}}
\end{table}

\begin{table}[!t]
\setlength{\abovecaptionskip}{-0.001in}
\renewcommand{\arraystretch}{1}
\caption{FROC (\%) of compared methods and our SGDA on target dataset tianchi w.r.t percentage of training images.}
\label{cross1_tianchi_table}
\centering
\setlength{\tabcolsep}{1.5mm}{
\begin{tabular}{l|c|c c c c c} \hline
    Method & tianchi & 20\% & 40\% & 60\% & 80\% & 100\% \\ \hline
    uniNoduleNet & 68.23 & 69.08 & 69.45 & 69.13 & 69.40 & 69.98 \\
    DANoduleNet \cite{wang2019towards} & - & 66.95 & 68.60 & 69.29 & 68.60 & 69.88 \\
    SGDANoduleNet & - & \textbf{73.61} & \textbf{73.61} & \textbf{74.84} & \textbf{75.00} & \textbf{74.57} \\ \hline
    \end{tabular}}
\end{table}

\begin{table}[!t]
\setlength{\abovecaptionskip}{-0.001in}
\renewcommand{\arraystretch}{1}
\caption{FROC (\%) of compared methods and our SGDA on target dataset russia w.r.t percentage of training images.}
\label{cross1_russia_table}
\centering
\setlength{\tabcolsep}{1.55mm}{
\begin{tabular}{l|c|c c c c c} \hline
    Method & russia & 20\% & 40\% & 60\% & 80\% & 100\% \\ \hline
    uniNoduleNet & 37.19 & 38.06 & 39.66 & 39.65 & 40.60 & 41.07 \\
    DANoduleNet \cite{wang2019towards} & - & 40.16 & 41.66 & 41.15 & \textbf{42.50} & \textbf{42.30} \\
    SGDANoduleNet & - & \textbf{40.20} & \textbf{41.78} & \textbf{41.59} & 42.10 & 41.43 \\ \hline
    \end{tabular}}
\end{table}

\begin{table}[!t]
\setlength{\abovecaptionskip}{-0.001in}
\renewcommand{\arraystretch}{1}
\caption{FROC (\%) of compared methods and our SGDA on tianchi and russia. The 1st and 2nd rows are FROCs of models trained and tested on two datasets, while the rest two rows are FROCs of models trained and tested on all three datasets. }
\label{cross1_tianchi_russia}
\centering
\setlength{\tabcolsep}{5mm}{
\begin{tabular}{l|c|c|c} \hline
    Method & tianchi & russia & Avg \\ \hline
    uniNoduleNet & 73.88 & 37.31 & 55.59 \\
    SGDANoduleNet & 74.57 & 37.75 & \underline{56.16} \\ \hline
    uniNoduleNet & 68.60 & 33.35 & 50.97 \\
    SGDANoduleNet & 77.13 & 37.15 & \textbf{57.14} \\ \hline
    \end{tabular}}
\end{table}

\subsubsection{Cross-center Experiments}

We first compare the performance of the NoduleNets \cite{10.1007/978-3-030-32226-7_30} trained and tested on the same datasets `self-test' with the models trained using other datasets, such as `tianchi-trained'. The results are presented in Fig.~\ref{small_selftest_hist}. It can be seen that the models trained using samples from other domains usually perform worse than the self-test models. This demonstrates the necessity of conducting adaptation cross different domains. An exception is that the self-test model on the russia \cite{russia} dataset is comparable to the models trained using other datasets. The main reason is that there are quite limited number of CT images and large quantity of small nodules in the russia \cite{russia} dataset. This makes it a very challenging dataset, and thus it is hard to obtain a well-performing self-test model. Then we conduct experiments to evaluate the proposed SGDA in dealing with cross-center \cite{DBLP:journals/tcbb/JiangGWHXQL21} pulmonary nodule detection using LUNA16 \cite{DBLP:journals/mia/SetioTBBBCCDFGG17}, tianchi \cite{tianchi}, and russia \cite{russia}.

\begin{table}[!t]
\setlength{\abovecaptionskip}{-0.001in}
\renewcommand{\arraystretch}{1}
\caption{FROC (\%) of compared methods and our SGDA on LUNA16 and russia.}
\label{cross1_luna16_russia}
\centering
\setlength{\tabcolsep}{4.8mm}{
\begin{tabular}{l|c|c|c} \hline
    Method & LUNA16 & russia & Avg \\ \hline
    uniNoduleNet & 79.94 & 37.98 & 58.96 \\
    SGDANoduleNet & 79.42 & 39.25 & \underline{59.33} \\ \hline
    uniNoduleNet & 79.88 & 33.35 & 56.61 \\
    SGDANoduleNet & 81.91 & 37.15 & \textbf{59.53} \\ \hline
    \end{tabular}}
\end{table}

\begin{table}[!t]
\setlength{\abovecaptionskip}{-0.001in}
\renewcommand{\arraystretch}{1}
\caption{FROC (\%) of compared methods and our SGDA on LUNA16 and tianchi.}
\label{cross1_luna16_tianchi}
\centering
\setlength{\tabcolsep}{4.6mm}{
\begin{tabular}{l|c|c|c} \hline
    Method & LUNA16 & tianchi & Avg \\ \hline
    uniNoduleNet & 81.06 & 69.08 & 75.07 \\
    SGDANoduleNet & 79.81 & 73.61 & \underline{76.71} \\ \hline
    uniNoduleNet & 79.88 & 68.60 & 74.24 \\
    SGDANoduleNet & 81.91 & 77.13 & \textbf{79.52} \\ \hline
    \end{tabular}}
\end{table}

\begin{table*}[!t]
\setlength{\abovecaptionskip}{-0.001in}
\renewcommand{\arraystretch}{1}
\caption{Comparison of our SGDA and other multi-domain methods in terms of FROC on dataset PN9. The values are pulmonary nodule detection sensitivities (unit: \%) with each column representing the average number of false positives per CT image. All the methods utilize SANet as backbone: (1) baseline model with the prefix `uni-', (2) universal models with `SG' in the name (Ours).}
\label{multi_PN9}
\centering
\setlength{\tabcolsep}{2.2mm}{
\begin{tabular}{l|c|c|c|c|c c c c c c c|c} \hline
    Method & \#Adapters & \#Groups & \#Params & \#FLOPs & 0.125 & 0.25 & 0.5 & 1.0 & 2.0 & 4.0 & 8.0 & Avg \\ \hline
    uniSANet \cite{mei2021sanet} & - & - & 15.28M & 138G & 38.08 & 45.05 & 54.46 & 64.50 & 75.33 & 83.86 & 89.96 & 64.46 \\
    DASANet \cite{wang2019towards} & 3 & - & 15.32M & 138G & 54.86 & 54.86 & 54.86 & 64.94 & 75.43 & 83.53 & 88.18 & 68.09 \\
    *SGDASANet w/o CA & 3 & 4 & 15.36M & 138G & 52.06 & 52.06 & \textbf{58.63} & \textbf{66.33} & \textbf{77.05} & \textbf{85.13} & \textbf{90.12} & 68.77 \\
    *SGDASANet w/ CA & 3 & 4 & 15.45M & 139G & \textbf{57.63} & \textbf{57.63} & 57.63 & 65.73 & 75.09 & 83.56 & 88.25 & \textbf{69.36} \\ \hline
    \end{tabular}}
\end{table*}

We mainly compare with the uniNoduleNet baseline, and the most competitive multi-domain counterpart DANoduleNet.
The models are trained on two of the three datasets, and then the remaining dataset is treated as the target dataset for the trained network to be finetuned and tested on. To better investigate the cross-center detection performance, we finetune the trained network using a varied percentage (20\%, 40\%, 60\%, 80\%, and 100\%) of training images in the target dataset, and test the finetuned network on the whole testing set. The results are shown in Table~\ref{cross1_luna16_table}, Table~\ref{cross1_tianchi_table}, and Table~\ref{cross1_russia_table}. We also list the FROCs of NoduleNets \cite{10.1007/978-3-030-32226-7_30}, which are trained using all the training images in the corresponding target datasets. We can see that our SGDA achieves the best performance in most cases, and improves the baseline by up to approximately 6\% in terms of the FROC score for domain generalization. The barely satisfactory results in Table~\ref{cross1_russia_table} may be due to the limited depth information resulting from the resampling operations on LUNA16 \cite{DBLP:journals/mia/SetioTBBBCCDFGG17} and tianchi \cite{tianchi}, and then the advantage of grouping is diminished. Besides, results of the models trained and tested on the corresponding two source datasets are reported in Table~\ref{cross1_tianchi_russia}, Table~\ref{cross1_luna16_russia}, and Table~\ref{cross1_luna16_tianchi}. The results indicate that uniNoduleNet trained on two datasets have better overall performance than the one trained using all the three datasets, while our SGDANoduleNet trained on three datasets performs better than the one trained using two datasets. This further indicates that existing pulmonary nodule detectors are not effective in handling multiple domains, while our SGDA can well exploit the information from multi-center datasets.

\begin{figure}[!t]
\centering
\setlength{\abovecaptionskip}{-0.01in}
\includegraphics[width=3.4in]{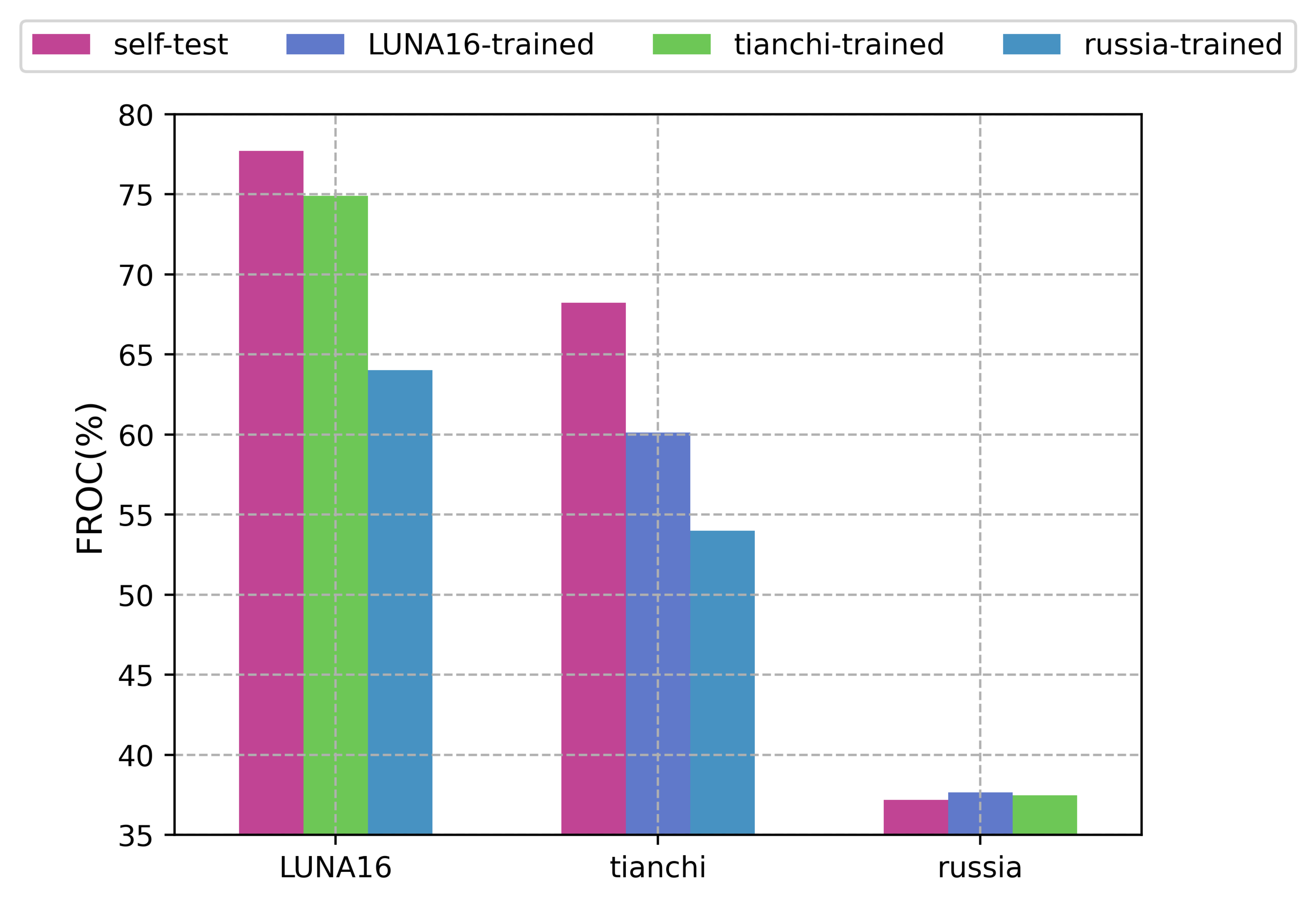}
\caption{Comparison of the self-test models and models trained using other datasets.}
\label{small_selftest_hist}
\end{figure}

\subsection{Large-Scale Experiments}
Our large-scale experiments are conducted on the PN9 \cite{mei2021sanet} dataset, and we adopt the SANet \cite{mei2021sanet} as the backbone, which is initialized with the weights pretrained on PN9.
The same data split as \cite{mei2021sanet} is adopted for the ease of comparison. We use the Stochastic Gradient Descent (SGD) optimizer with a batch size of 16. The learning rate of our SGDA module is 0.01, and the whole network except the SGDA module is set to be 0.001; the momentum and weight decay coefficients are set to be 0.9 and $1\times10^{-4}$, respectively. We use the same hyperparameters as \cite{mei2021sanet} for all networks except for the number of epochs of training without RCNN. It is tuned for each network to obtain the best performance. All the large-scale experiments are implemented using PyTorch on 2 NVIDIA GeForce RTX 3090 GPU with 24GB memory.


\subsubsection{Multi-center Experiments}
\label{large_multi}

The PN9 \cite{mei2021sanet} dataset is naturally a multi-center dataset since it is very large and contain diverse types of pulmonary nodules from different centers.
In this set of experiments, we mainly compare with:
(1) the uniSANet baseline model, which treats data from different centers equally; and (2) the most competitive DASANet, which is also a universal model and can be regarded as a special case of our SGDASANet, where `w/o CA' and `w/ CA' represent our models without and together with the three way cross attention strategy. `*' signifies that our SGDA module is added in each 3D residual block of the SANet \cite{mei2021sanet} backbone except for the first three blocks. As reported in Table~\ref{multi_PN9}, our SGDA is superior to other approaches. The FROC score is 69.36\%, which improves the baseline \cite{mei2021sanet} by 4.9\%.
This further demonstrates the effectiveness of our proposed SGDA. Besides, as seen in Table~\ref{experimental_results} and Table~\ref{multi_PN9}, the extra computational cost from our slice grouped domain attention (SGDA) is acceptable especially when the grouping trick is adopted in the three-way cross attention. This may be because that the computational cost of our proposed SGSE adapter and domain assignment component is neglectable compared with that of the 3D convolutions. After the grouping trick is applied in our proposed three-way cross attention, its computational burden, which mainly comes from the matrix multiplication, is largely reduced and becomes more acceptable.

\begin{figure}[!t]
\centering
\setlength{\abovecaptionskip}{-0.01in}
\includegraphics[width=3in]{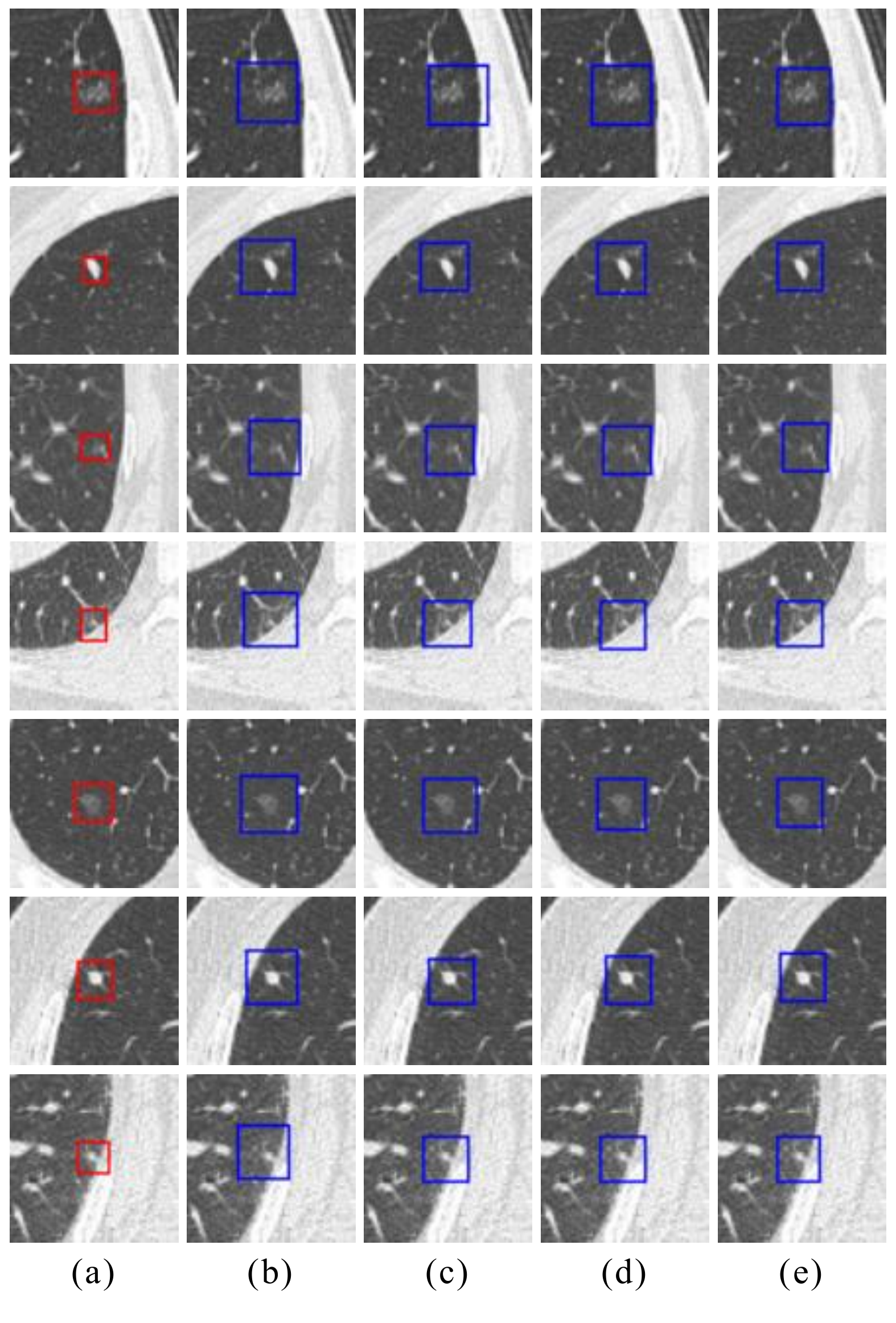}
\caption{Qualitative comparison of central slices for our SGDA and other multi-domain methods on the large-scale multi-center dataset PN9. (a) Ground truth. (b) Baseline \cite{mei2021sanet}. (c)-(e) Detection results of pulmonary nodule detection networks + DA, SGDA w/o CA and SGDA w/ CA.}
\label{large_visualization}
\end{figure}

The visualization of central slices for nodule ground truths and the above four models' detection results is shown in Fig.~\ref{large_visualization}. It can be seen that the detected nodule positions of our *SGDASANet are more accurate and with smaller bounding boxes.

\begin{table}[!t]
\setlength{\abovecaptionskip}{-0.001in}
\renewcommand{\arraystretch}{1}
\caption{FROC (\%) of SANet trained using PN9 and tested on the PN9, LUNA16, tianchi and russia datasets.}
\label{large_selftest_table}
\centering
\setlength{\tabcolsep}{4mm}{
\begin{tabular}{l|c c c c} \hline
     & PN9 & LUNA16 & tianchi & russia \\ \hline
    SANet \cite{mei2021sanet} & 64.46 & 59.56 & 58.84 & 20.77 \\ \hline
    \end{tabular}}
\end{table}

\begin{figure}[!t]
\centering
\setlength{\abovecaptionskip}{-0.01in}
\includegraphics[width=3in]{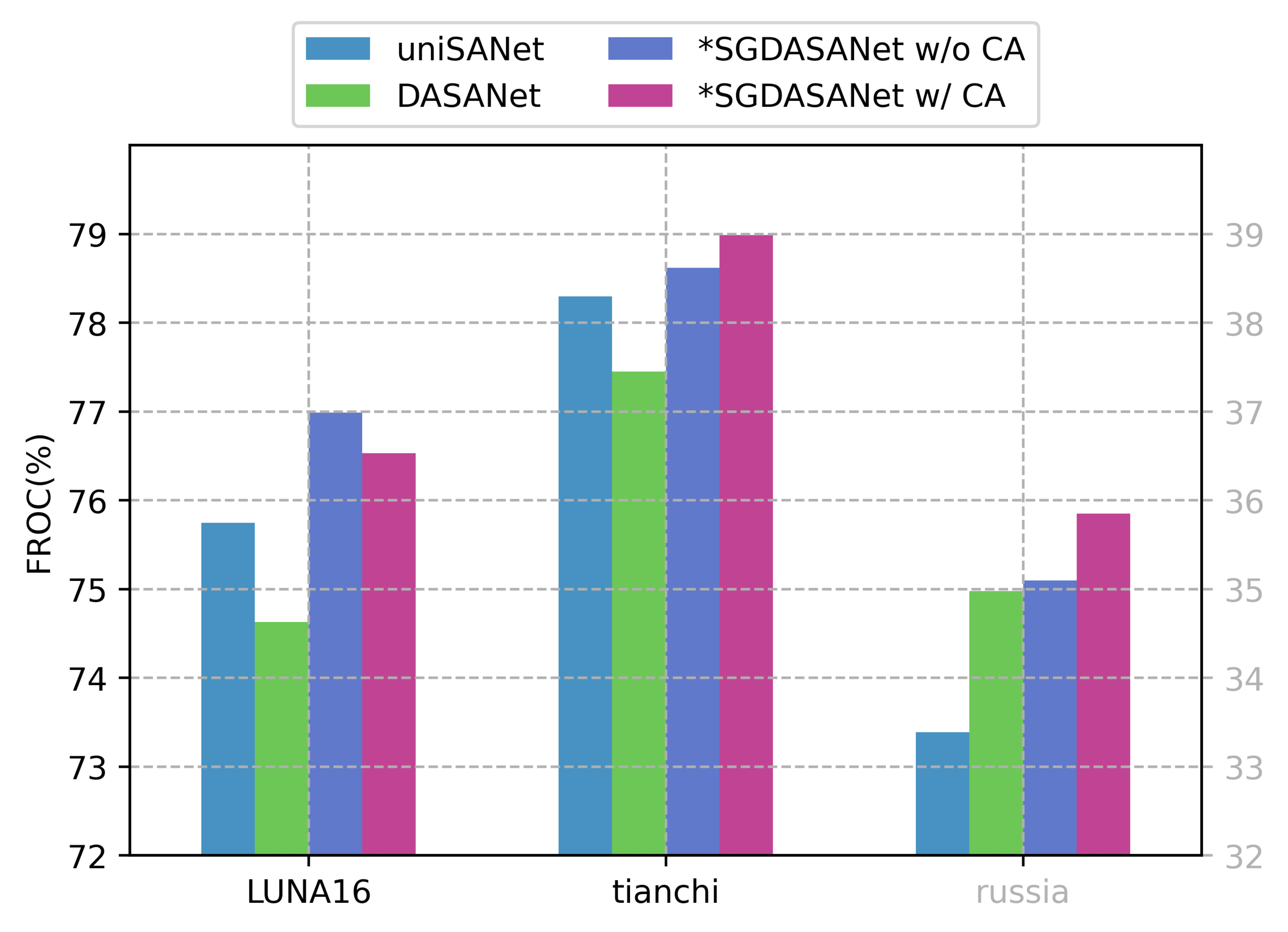}
\caption{FROC (\%) of compared methods and our SGDA on LUNA16, tianchi, and russia w.r.t 20\% of training images. The scale of LUNA16 and tianchi is on the left, while the scale of russia is on the right.}
\label{cross2_hist}
\end{figure}

\subsubsection{Cross-center Experiments}
We first train SANet \cite{mei2021sanet} using PN9 \cite{mei2021sanet} and test the obtained model on PN9 \cite{mei2021sanet}, LUNA16 \cite{DBLP:journals/mia/SetioTBBBCCDFGG17}, tianchi \cite{tianchi}, and russia \cite{russia}, respectively. The results reported in Table~\ref{large_selftest_table} illustrate that a model trained using the source dataset may perform well on the source dataset but suffer from performance degradation when tested on a different domain. Then we perform large-scale experiments to evaluate the proposed SGDA in dealing with cross-center pulmonary nodule detection. We take PN9 \cite{mei2021sanet} as the source dataset, and the LUNA16 \cite{DBLP:journals/mia/SetioTBBBCCDFGG17}, tianchi \cite{tianchi}, and russia \cite{russia} as the target dataset in turn. Each model trained on PN9 \cite{mei2021sanet} from Sec. \ref{large_multi} is finetuned on the latter three datasets separately using 20\% of the training images, and tested on the whole testing set. The results are shown in Fig. \ref{cross2_hist}, which demonstrates the effectiveness of our SGDA for improving the model generalization ability.

\section{Conclusion}
\label{sec:Conclusion}

In this paper, we study the challenging problem of universal pulmonary nodule detection. We propose a slice grouped domain attention (SGDA) module, which aims to enhance the generalization capability of the pulmonary nodule detection networks. It is a \textit{universal} plug-and-play module, which can be incorporated into existing backbone networks, and works on multiple pulmonary nodule datasets with no requirement for prior domain knowledge. Extensive experimental results show that schemes powered by SGDA achieve the state-of-the-art performance in both multi-center and cross-center pulmonary nodule tasks. In the future, we intend to apply dynamic networks to reduce computational cost, and verify effectiveness of our method on more large-scale datasets.

\section*{Acknowledgment}
The authors would like to thank the handling associate editor and all the anonymous reviewers for their constructive comments. This research was supported in part by the National Key Research and Development Program of China under No. 2021YFC3300200,
the Special Fund of Hubei Luojia Laboratory under Grant 220100014, and the National Natural Science Foundation of China (Grant No. 62276195 and 62141112).


\ifCLASSOPTIONcaptionsoff
  \newpage
\fi



\bibliographystyle{IEEEtran}
\bibliography{TCBBSI_SGDA}
%



%

\begin{IEEEbiography}[{\includegraphics[width=1in,height=1.25in,clip,keepaspectratio]{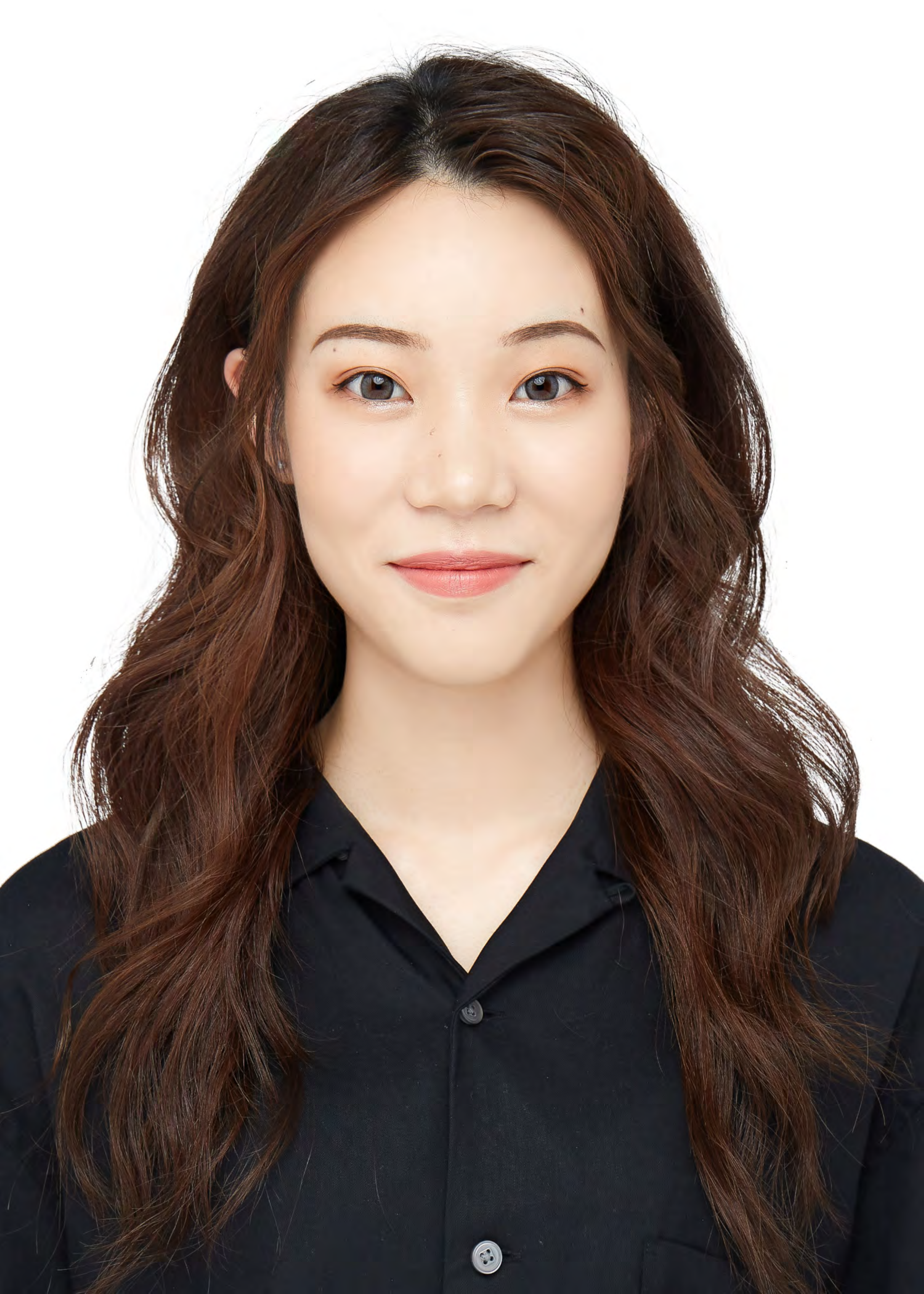}}]{Rui Xu}
is a Ph.D. student in the School of Computer Science, Wuhan University, Wuhan, China. She is supervised by Prof. Yong Luo and Prof. Bo Du. She received her bachelor's degree from Jilin University, Changchun, China, in 2018, and her master's degree from the University of California, Los Angeles, CA, US, in 2020. Her research interests include computer vision and machine learning.
\end{IEEEbiography}

\begin{IEEEbiography}[{\includegraphics[width=1in,height=1.25in,clip,keepaspectratio]{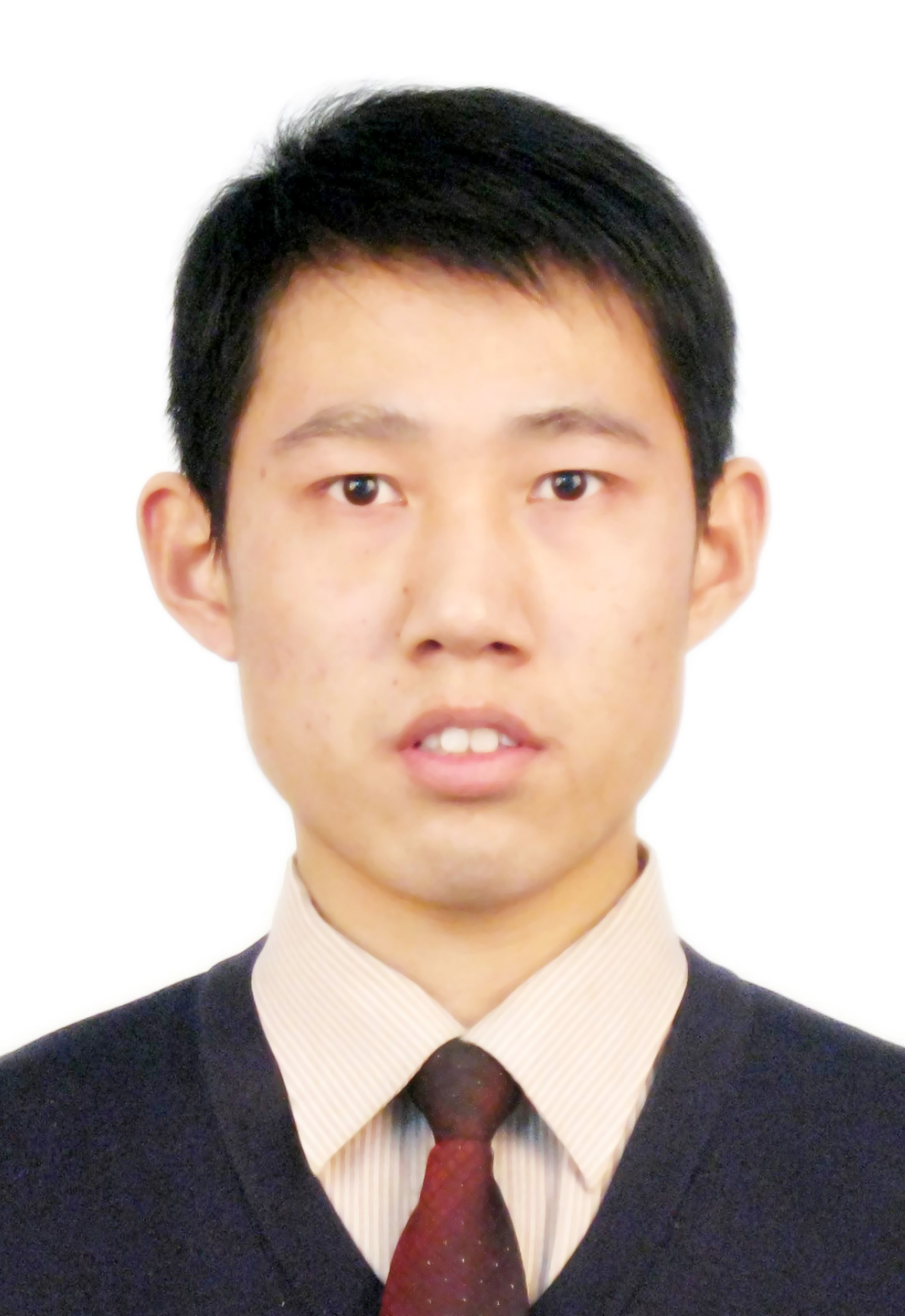}}]{Zhi Liu}
(S'11-M'14-SM'19) received his Ph.D. degree in informatics in National Institute of Informatics. He is currently an Associate Professor at The University of Electro-Communications. His research interest includes video network transmission and mobile edge computing. He is now an editorial board member of Springer wireless networks and IEEE Open Journal of the Computer Society. He is a senior member of IEEE.
\end{IEEEbiography}

\begin{IEEEbiography}[{\includegraphics[width=1in,height=1.25in,clip,keepaspectratio]{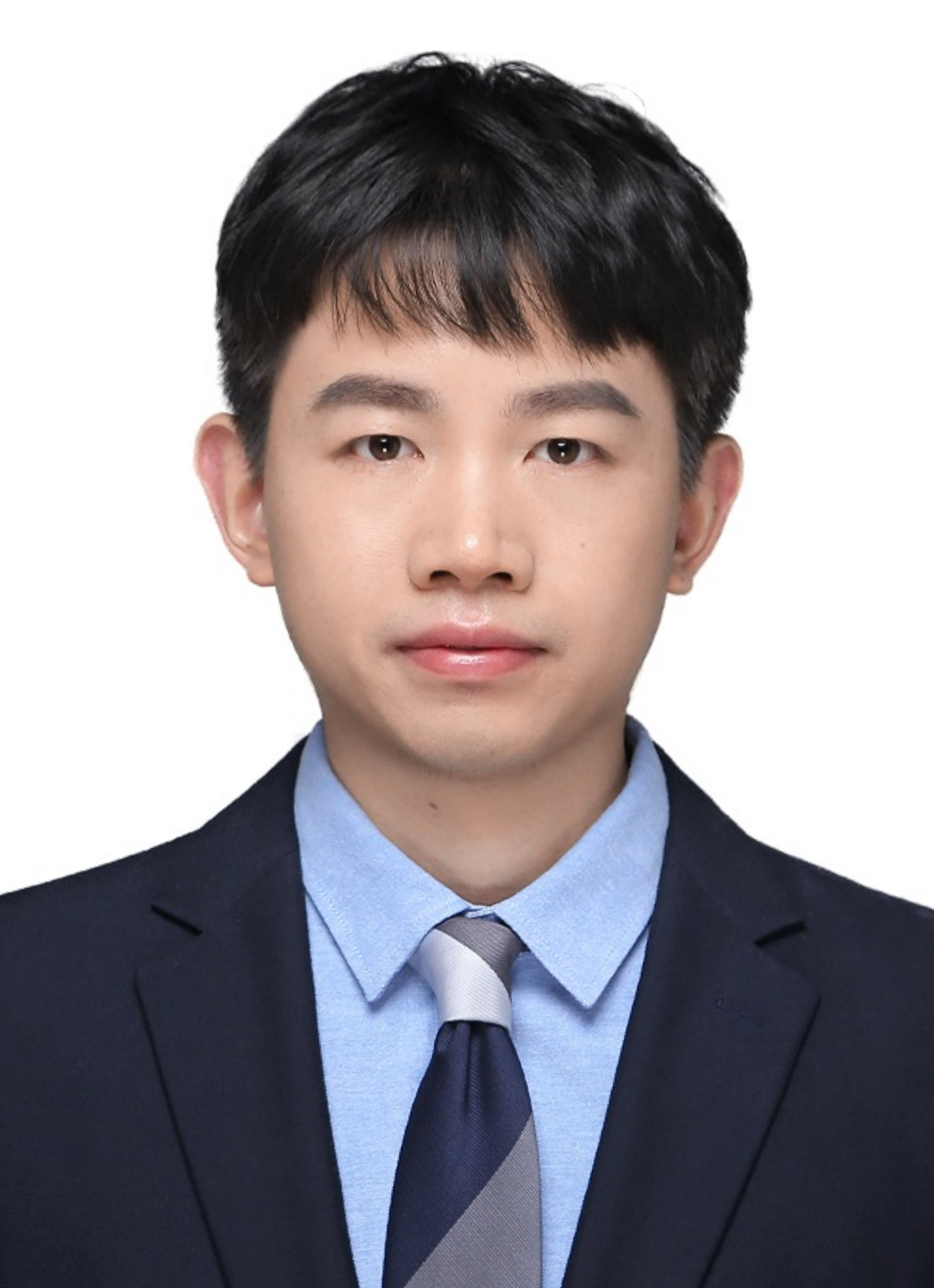}}]{Yong Luo}
received the B.E. degree in Computer Science from the Northwestern Polytechnical University, Xi'an, China, and the D.Sc. degree in the School of Electronics Engineering and Computer Science, Peking University, Beijing, China. He is currently a Professor with the School of Computer Science, Wuhan University, China. His research interests are primarily on machine learning and data mining with applications to visual information understanding and analysis. He has authored or co-authored over 60 papers in top journals and prestigious conferences including IEEE T-PAMI, IEEE T-NNLS, IEEE T-IP, IEEE T-KDE, IEEE T-MM, ICCV, WWW, IJCAI and AAAI. He is serving on editorial board for IEEE T-MM. He received the IEEE Globecom 2016 Best Paper Award, and was nominated as the IJCAI 2017 Distinguished Best Paper Award. He is also a co-recipient of the IEEE ICME 2019 and IEEE VCIP 2019 Best Paper Awards.
\end{IEEEbiography}

\begin{IEEEbiography}[{\includegraphics[width=1in,height=1.25in,clip,keepaspectratio]{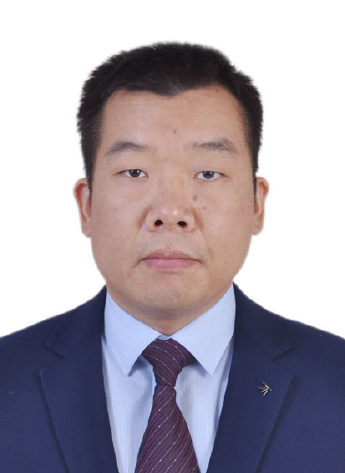}}]{Han Hu}
received the B.E. and Ph.D. degrees from the University of Science and Technology of China, China, in 2007 and 2012, respectively. He is currently a Professor with the School of Information and Electronics, Beijing Institute of Technology, China. His research interests include multimedia networking, edge intelligence, and space-air-ground integrated network. He received several academic awards, including the Best Paper Award of the IEEE TCSVT 2019, the Best Paper Award of the IEEE Multimedia Magazine 2015, and the Best Paper Award of the IEEE Globecom 2013. He has served as an Associate Editor of IEEE TMM and Ad Hoc Networks, and a TPC member of Infocom, ACM MM, AAAI, and IJCAI.
\end{IEEEbiography}

\begin{IEEEbiography}[{\includegraphics[width=1in,height=1.25in,clip,keepaspectratio]{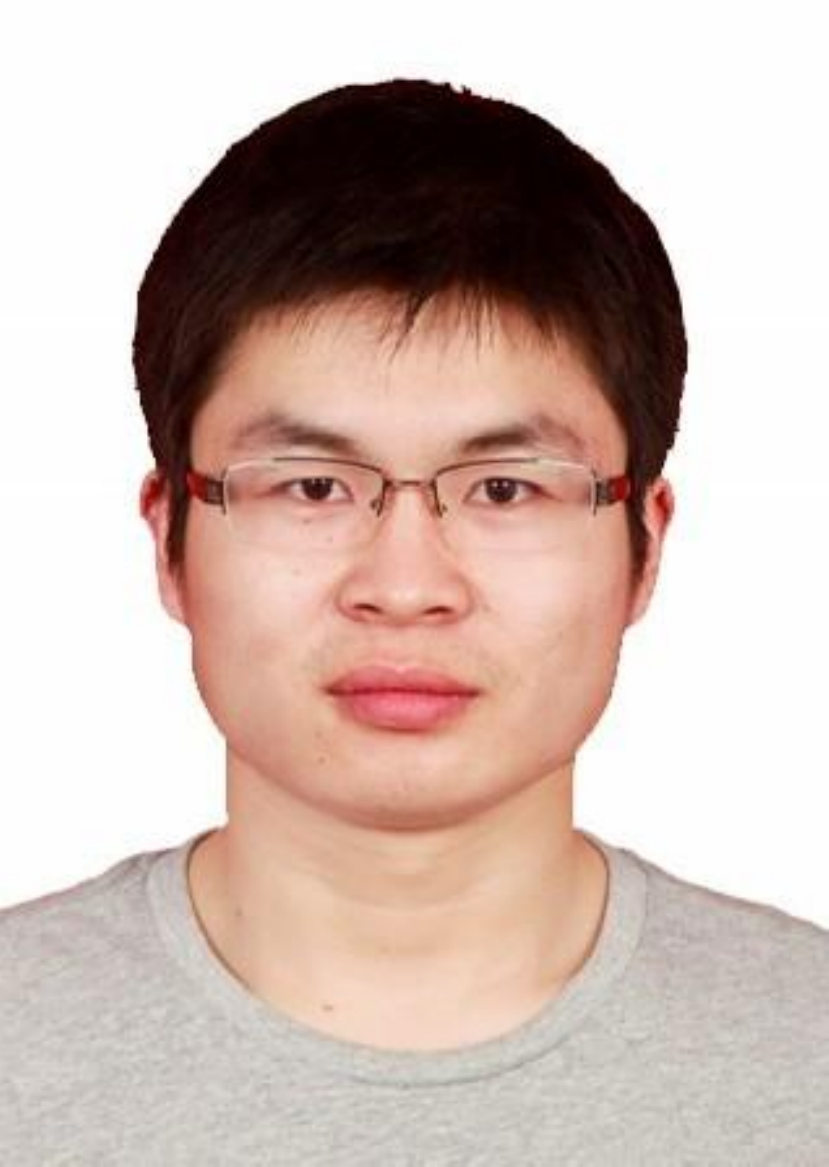}}]{Li Shen}
received his Ph.D. in school of mathematics, South China University of Technology in 2017. He is currently a research scientist at JD Explore Academy, China. Previously, he was a research scientist at Tencent AI Lab, China. His research interests include theory and algorithms for large scale convex/nonconvex/minimax optimization problems, and their applications in statistical machine learning, deep learning, reinforcement learning, and game theory.
\end{IEEEbiography}

\begin{IEEEbiography}[{\includegraphics[width=1in,height=1.25in,clip,keepaspectratio]{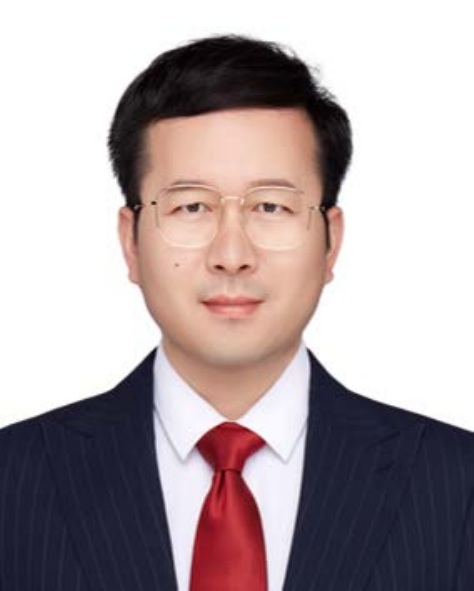}}]{Bo Du}
received the PhD degree from the State Key Laboratory of Information Engineering in Surveying, Mapping and Remote Sensing, Wuhan University, Wuhan, China, in 2010. He is currently a professor with the School of Computer, Wuhan University. He has published more than 100 scientific papers, such as the IEEE Transactions on Geoscience and Remote Sensing, IEEE Transactions on Neural Networks and Learning Systems, IEEE Transactions on Image Processing, IEEE Transactions on Cybernetics, AAAI, and IJCAI. His research interests include pattern recognition, hyperspectral image processing, and signal processing.
\end{IEEEbiography}

\begin{IEEEbiography}[{\includegraphics[width=1in,height=1.25in,clip,keepaspectratio]{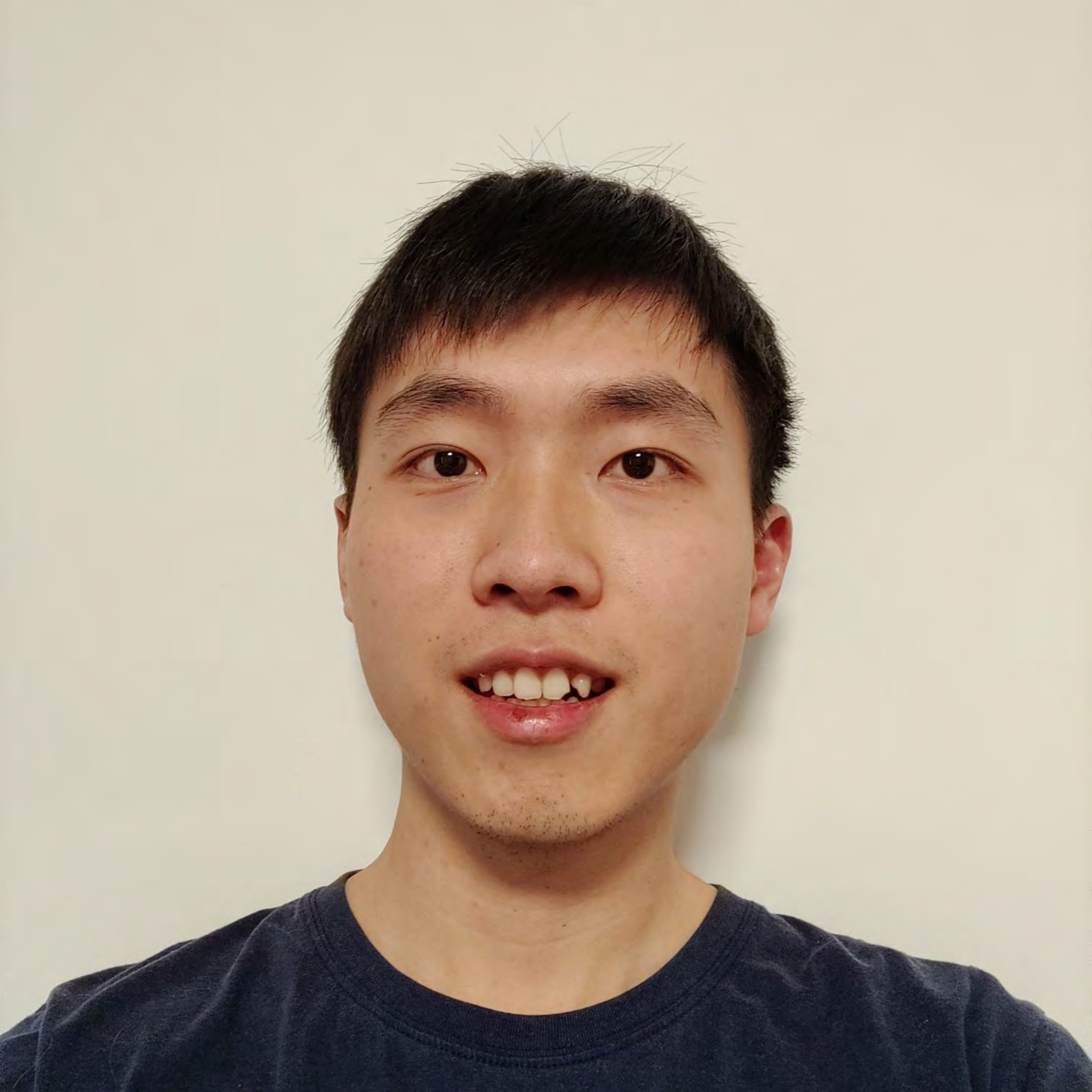}}]{Kaiming Kuang}
is currently an Algorithm Engineer at Dianei Technology Co., Ltd. in Shanghai, China. He earned his B.Sc. degree at Renmin University of China, Beijing, China, and received M.Sc. degree at University of California San Diego, La Jolla, US. His research focuses in medical AI, with emphasis on medical computer vision. He has published several papers in MICCAI, EbioMedicine and Frontiers in Oncology.
\end{IEEEbiography}

\begin{IEEEbiography}[{\includegraphics[width=1in,height=1.25in,clip,keepaspectratio]{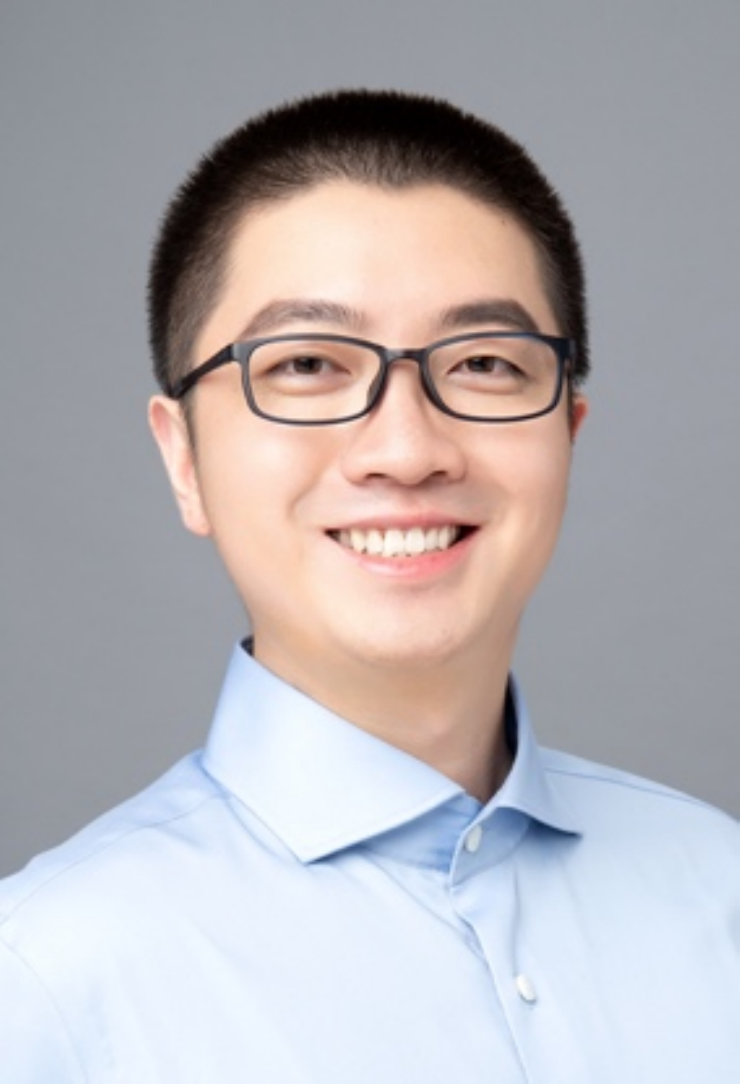}}]{Jiancheng Yang}
received the B.Eng. and M.Eng. degree in Automation from Shanghai Jiao Tong University, China, in 2015 and 2018, respectively. He also received the engineer degree (double master's degree) in Institut Mines-Telecom, France, in 2016. He is working towards the Ph.D. degree with Shanghai Jiao Tong University. He was a visiting researcher at Harvard University, USA and EPFL, Switzerland. His research interests center around the interdisciplinary field of medical image analysis and 3D computer vision.
\end{IEEEbiography}







\end{document}